\documentclass{emulateapj}
\usepackage[]{natbib}
\usepackage{graphicx, amsmath}
\bibpunct{(}{)}{,}{a}{}{,}

\begin{document}
\def\bullet{\object{1E0657$-$56}}
\def\bbullet{\object{MACS~J0025.4$-$1222}}
\def\HST{{\it HST}}

\def\arcsecf{\!\!^{\prime\prime}}
\def\arcminf{\!\!^{\prime}}
\def\diff{\mathrm{d}}
\def\ngx{N_{\mathrm{x}}}
\def\ngy{N_{\mathrm{y}}}
\def\eck#1{\left\lbrack #1 \right\rbrack}
\def\eckk#1{\bigl[ #1 \bigr]}
\def\round#1{\left( #1 \right)}
\def\abs#1{\left\vert #1 \right\vert}
\def\wave#1{\left\lbrace #1 \right\rbrace}
\def\ave#1{\left\langle #1 \right\rangle}
\def\kms{{\rm \:km\:s}^{-1}}
\def\dds{D_{\mathrm{ds}}}
\def\dd{D_{\mathrm{d}}}
\def\ds{D_{\mathrm{s}}}
\def\cs{\mbox{cm}^2\mbox{g}^{-1}}
\def\magz{m_{\rm z}}

\title{Focusing Cosmic Telescopes:
Exploring Redshift $z\sim 5-6$ Galaxies with
the Bullet Cluster {\bullet}\altaffilmark{*}}
\altaffiltext{*}{Based on
observations made with the NASA/ESA Hubble Space Telescope, obtained
at the Space Telescope Science Institute, which is operated by the
Association of Universities for Research in Astronomy, Inc., under
NASA contract NAS 5-26555. These observations are associated with
programs \# GO10200, GO10863, and GO11099.}
\shorttitle{}
\author{Maru\v{s}a Brada\v{c}\altaffilmark{1,2,x},
Tommaso Treu\altaffilmark{1,y},
Douglas Applegate\altaffilmark{3},
Anthony \ H. Gonzalez\altaffilmark{4},
Douglas \ Clowe\altaffilmark{5},
William \ Forman\altaffilmark{6},
Christine\ Jones\altaffilmark{6},
Phil Marshall\altaffilmark{1},
Peter Schneider\altaffilmark{7},
Dennis\ Zaritsky\altaffilmark{8}}
\shortauthors{Brada\v{c} et al.}
\altaffiltext{1}{Department of Physics, University of California, Santa Barbara, CA 93106, USA}
\altaffiltext{2}{Department of Physics, University of California, Davis, CA 95616, USA}
\altaffiltext{3}{Kavli Institute for Particle Astrophysics and
  Cosmology, Stanford University, 382 Via Pueblo Mall, Stanford, CA
  94305-4060, USA}
\altaffiltext{4}{Department of Astronomy, University of Florida, 211 Bryant Space Science Center, Gainesville, FL 32611, USA}
\altaffiltext{5}{Department of Physics \& Astronomy, Ohio University, Clippinger Labs 251B, Athens, OH 45701
}
\altaffiltext{6}{Harvard-Smithsonian Center for Astrophysics, 60 Garden Street, Cambridge, MA 02138, USA}
\altaffiltext{7}{Argelander-Institut f\"{u}r Astronomie, Auf dem
    H\"{u}gel 71, D-53121 Bonn, Germany}
\altaffiltext{8}{Steward Observatory, University of Arizona, 933 N Cherry Ave., Tucson, AZ 85721, USA}
\altaffiltext{x}{Hubble Fellow}
\altaffiltext{y}{Sloan Fellow, Packard Fellow}
\email{marusa@physics.ucdavis.edu}


\begin{abstract}
  The gravitational potential of clusters of galaxies acts as a cosmic
  telescope allowing us to find and study galaxies at fainter limits
  than otherwise possible and thus probe closer to the epoch of
  formation of the first galaxies.  We use the Bullet Cluster
  {\bullet} ($z = 0.296$) as a case study, because its high mass and
  merging configuration makes it one of the most efficient cosmic
  telescopes we know.  We develop a new algorithm to reconstruct the
  gravitational potential of the Bullet Cluster\, based on a
  non-uniform adaptive grid, combining strong and weak gravitational
  lensing data derived from deep \HST/ACS F606W-F775W-F850LP and
  ground-based imaging.  We exploit this improved mass map to study
  $z\sim 5-6$ Lyman Break Galaxies (LBGs), which we detect as
  dropouts. One of the LBGs is multiply imaged, providing a geometric
  confirmation of its high redshift, and is used to further improve
  our mass model. We quantify the uncertainties in the magnification
  map reconstruction in the intrinsic source luminosity, and in the
  volume surveyed, and show that they are negligible compared to
  sample variance when determining the luminosity function of
  high-redshift galaxies.  With shallower and comparable magnitude
  limits to HUDF and GOODS, the Bullet cluster observations, after
  correcting for magnification,  probe deeper into the luminosity
    function of the high redshift galaxies than GOODS and only
    slightly shallower than HUDF. We conclude that accurately focused
  cosmic telescopes are the most efficient way to sample the bright
  end of the luminosity function of high redshift galaxies and - in
  case they are multiply imaged - confirm their redshifts.
\end{abstract}


\section{Introduction}
\label{sec:intro}

Studying galaxies at high redshifts is crucial for understanding both
their formation and evolution, and the role they played in cosmic
reionization.  However, recent observations of $z\gtrsim 6$ objects
(e.g., \citealp{bouwens08}) show that UV-bright galaxies alone are not
believed to be sufficient to reionize the Universe (unless for
example, the escape fractions or clumping factors are dramatically
different from current assumptions). It is therefore important to
continue to find radiation sources at the highest redshifts and study
their physical properties.

Observations of galaxies at these high redshifts are challenging, not
only due to the large luminosity distance to these objects, but also
due to their lower luminosity, since the luminosity function evolves
compared to galaxies at moderate redshifts ($z\sim2$; see
e.g. \citealp{reddy09,bouwens07}).  Recently our knowledge of the
early cosmic epoch of galaxy formation ($z\gtrsim 4$) has vastly
increased, providing an opportunity for us to test theories of star
formation \citep[and references therein]{stark09,bouwens08, bouwens07,
  yoshida06, stanway05, giavalisco04}. Comparing the measured
luminosity function with results from simulations provides insight on
the efficiency of star formation as a function of mass scale, and on
the feedback mechanisms that may limit it (see
e.g. \citealp{vandenbosch03}).

We can find high-redshift galaxies by searching for the redshifted
Lyman break using broad band photometry \citep{steidel96}.  These
sources, known as Lyman Break Galaxies (LBGs, see
e.g. \citealp{vanzella09} for recent studies, and
\citealp{giavalisco02} for a review) are the best studied and the
largest sample of galaxies at redshifts $z\gtrsim 5$.  One can
identify $z\simeq 5$ objects by their non-detection in the $V$-band
and blueward (e.g. \citealp{giavalisco04b,stanway05,hildebrandt09,
  stark09}): such objects are referred to as ``$V$-band dropouts.''
Similarly, objects at $z\simeq 6$ are associated with $i$-band
non-detection (e.g. \citealp{stanway03, bouwens08}), and $z\simeq
7$\nobreakdash--$8$ with $z$-band (e.g. \citealp{bouwens06,bouwens08,
  henry09, oesch09}). To date, preliminary results using the new
WFC3/IR data have now been released. \citet{oesch09b} and
\citet{bunker09} also study the $z$-band dropouts, whereas
\citet{bouwens09b}, \citet{yan09}, and \citet{mclure09} study $Y$-band
dropouts which are potential $z\sim 8-8.5$ objects. At even higher
redshifts, $z\gtrsim 8$ objects are observed as $J$-band dropouts
(e.g. \citealp{yan09}, \citealp{henry08}, and
\citealp{bouwens05}).\footnote{The $z\sim9$ candidate object from
  \protect\citealp{henry08} was later confirmed to be a lower redshift
  one and the number of candidates from \citealp{bouwens05} appears
  consistent with the expected contamination from low-redshift
  interlopers.} The main limitations of these experiments to date are
the relatively low expected number counts (even in the deepest
fields), and contamination ($z \sim 2$ dusty galaxies and L and T
dwarf stars can also share the characteristic dropout appearance).
Furthermore, the depths probed by these surveys are around
characteristic luminosity for high-redshift population $L_{*}$ and do
not reach far into the low luminosity regime where the faint-end slope
is constrained, especially at higher redshifts.

We argue in this paper that at the luminosity regime down to $L\sim
L_{*}$, determining the high redshift luminosity function can be
performed more efficiently when using galaxy clusters as cosmic
telescopes with respect to blank fields. This technique was proposed
shortly after the first gravitationally-lensed arcs in galaxy clusters
were discovered \citep{soucail90} and has been successfully applied to
ever-improving data (e.g.,
\citealp{zheng09,bouwens09,richard06,ellis01}). Observations of galaxy
clusters used as cosmic telescope is consistently
delivering record holders in the search for the highest redshift
galaxies \citep{kneib04,bradley08}. The reason for this is that the
magnifying power of a massive cluster lens typically allows the
detection of objects more than a magnitude fainter than the
observation limit. Due to this same magnification, the effective solid
angle of the survey volume decreases; however, since the luminosity
function is practically exponential at the magnitudes we probe, the
lensing magnification gives us a substantial net gain. Indeed, if the
faint end slope of the luminosity function at high redshift is
steep (e.g. \citealp{bouwens07}), we gain at magnitudes below and
around the characteristic magnitude $M^*$. One concern is that we need
to properly account for the errors introduced when we determine the
true observed volume with a cosmic telescope
(e.g. \citealp{bouwens09}). However, we quantify the uncertainties in
the magnification map and show that they are negligible compared to
sample variance when determining the luminosity function of
high-redshift galaxies.

Cosmic telescopes offer two further advantages.  First, the detected
sources behind a cluster lens are magnified in apparent size. As a
result, we can resolve smaller physical scales than would otherwise be
possible, and begin to actually measure the properties of $z\gtrsim6$
galaxies on a case-by-case basis. Resolving the galaxies also helps in
rejecting contamination by foreground cold stars. Second, some of the
sources will be multiply imaged. If this is the case, then they can be
readily distinguished from the main contaminants, typically
intermediate redshift galaxies and cold stars. The positions, where
multiply imaged systems form are redshift dependent, so a $z>5$
multiply imaged source can be readily distinguished from
(multiply-imaged) $z\sim 2$ source and singly imaged stars.  Thus, in
the magnitude range we observe here, a cosmic telescope survey is more
efficient than a competing imaging survey in unlensed fields to the
same depth.

The most effective gravitational lenses for such studies are massive
and highly elongated (in projection, on the plane of the sky) galaxy
clusters. It is easy to understand the first requirement, as more
massive objects are more powerful lenses. The ellipticity is important
as well, as the area of highest magnification is located at the semi
major-axis intersection with the critical curve (i.e. curve connecting points
of formally infinite magnification).  Equivalently,
sources close to the major axis cusp have the largest
magnification. The cluster of galaxies {\bullet}, discovered by
\citet{tucker95}, is one of the hottest, most X-ray luminous clusters
known. It is also a plane-of-the sky merger
(e.g. \citealp{markevitch02,barrena02}), hence giving a highly
elongated mass distribution in projection.

It is also at a favorable redshift, which contributes to make it a
very efficient lens. The critical curve radius decreases with
increasing lens redshift for a source at a fixed (higher) redshift.
One might thus think that lower redshift clusters $0.05< z \lesssim
0.2$ are better suited for these surveys. However, a small solid angle
coverage by cluster member galaxies is desirable, hence reducing the
effectiveness of clusters that are at too low redshift. Whereas
lensing efficiency scales with angular diameter distances between
observer, source, and the lens, the size of cluster members only
depends upon the latter. There is an optimal balance between the size
of the critical curve and the size of the cluster members. In
addition, a good cosmic telescope candidate should also show many
strongly lensed images (i.e. efficient also for lensing of $z\sim 1-3$
sources), which are needed to reliably reconstruct its magnification
map. Bullet Cluster, with $>10$ definite strongly lensed systems and
the lens redshift of $z\sim 0.3$,  is well suited for these
  studies.  Specifically, since we are trying to detect objects close
  to the detection threshold, we do not want to rely on perfect image
  subtraction of the cluster members.  For clusters at much higher
  redshifts, on the other hand, the critical curves (and the area of
  high magnification) are not large enough.

A key requirement for a cosmic telescope survey is an absolutely
calibrated, high-resolution mass reconstruction for the lens. This is
necessary to accurately convert the observed solid angle into the
actual cosmic volume, as well as to reconstruct the source fluxes and
so determine their luminosity. In this paper we will describe a newly
developed technique to reconstruct the mass distributions of galaxy
clusters.  As we will show, the technique allows us to improve the
accuracy in the very center of this cluster, which is crucial to
accurately predict the magnifications of dropouts (and other objects
in the field). The key feature of this new technique is the optimal
use of information from both multiple image systems (strong lensing
from deep high-resolution \HST/ACS data) and from the distortions of
the singly-imaged background sources (weak lensing, from combined ACS
and ground based data) using adaptive grid. Multi-resolution methods
have been used in either strong or weak gravitational lensing in the
past (see
e.g. \citealp{marshall06,diego07,merten08,deb08,jullo09}). The new
method described here, uses information from both strong and weak
lensing regimes, on a multi-level grid that allows us to efficiently
reconstruct features at the maximum resolution allowed by the
data. This new method and map supersedes our previously-inferred
potential map, which was limited in its resolution due to the
inability of the mass-modeling routine to simultaneously reconstruct
large angular scale features (constrained by the relatively low
signal-to-noise weak lensing data), and the smaller-scale features in
the high signal-to-noise strong lensing regime.

This paper is structured as follows. Section~\ref{sec:swunitedamr}
describes the improved method to reconstruct the mass distribution on
a non-uniform grid of pixels. Section~\ref{sec:data} describes the
data and its reduction procedures, and in Section~\ref{sec:results} we
present the mass reconstruction of Bullet
Cluster. Section~\ref{sec:dropouts} then gives the results of our
search for $z\sim 5-6$ objects using our newly-calibrated lens, and
discusses further the advantages and disadvantages of performing such
studies behind clusters.  Our conclusions are summarized in
Section~\ref{sec:conclusions}.

Throughout the paper we assume a $\Lambda$CDM cosmology with
$\Omega_{\rm m} = 0.3$, $\Omega_{\Lambda} = 0.7$, and Hubble constant
$H_0 = 70 {\rm \ km \: s^{-1}\:\mbox{Mpc}^{-1}}$. At the cluster
redshift $z_{\rm cl}=0.296$, the physical length scale is 4.4
kpc/arcsec. All the coordinates in this paper are given for the epoch
J2000.0. All magnitudes are given in the AB system.


\section{Strong and weak lensing mass reconstruction on a 
non-uniform adapted grid}
\label{sec:swunitedamr}

Our combined strong and weak lensing analysis follows the algorithm
first proposed by \citet{bartelmann96}. The implementation that
includes strong lensing is described in \citet{bradac04a} and
implemented on ACS data in \citet{bradac06, bradac08b,
  bradac08}. However, in these works the method showed limitations due
to the fact that the resolution of the reconstruction was uniform
across the observed field. Therefore, all our reconstructions were
limited to small fields (a few arcmin on a side).  By using a
reconstruction grid whose pixel scale varies across the field, we are
able to overcome these limitations and map the gravitational potential
of a cluster lens over a much larger field of view (Brada\v{c} et
al. 2009, in preparation). More importantly for this paper, we are
able to achieve increased resolution in the cluster centre (close to
where we see strongly lensed images), and hence the magnification map
in the regions of high magnification is more accurate.

All lensing observable quantities can be written as spatial
derivatives of the lens potential (see e.g. \citealp{bartelmann00}).
In the past we have modeled a cluster's projected gravitational
potential by a set of values on a regular grid $\psi_k =
\psi(\vec\theta^k)$, from which all predicted observables are then
evaluated by finite differencing (see e.g. \citealp{abramowitz}). For
example, the scaled surface mass density $\kappa$ is related to $\psi$
via the Poisson equation, $2\kappa = \nabla^2\psi$ (where the physical
surface mass density is $\Sigma = \kappa \: \Sigma_{\rm crit}$ and
$\Sigma_{\rm crit}$ depends upon the angular diameter distances
between the observer, the lens, and the source). Similarly, the shear
$\gamma = \gamma_1 + {\rm i} \gamma_2$ and the deflection angle $\vec
\alpha$  are derivatives of the potential: $\gamma_1 =
0.5(\psi_{,11}-\psi_{,22})$; $\gamma_2 = \psi_{,12}$; and $\vec \alpha
= \nabla \psi$.

In our improved method presented here, the regularity of the grid is relaxed. The
advantage of working on an irregular and adaptive grid is that it is
very flexible, which is crucial when dealing with merging clusters. As
a result we cannot use the standard finite differencing formulae of
\citet{abramowitz}, but instead extend them to allow for non-uniform
differences in the spatial variables (see also \citealp{deb08} for an
alternative approach).  We write down the Taylor expansion around the
point $\vec\theta^0$ where we want to estimate the lensing quantities,
in terms of $N$ neighbouring grid point locations $\vec\theta^{k}$
(note that $\vec\theta^0$ does not have to be one of the grid points).
This gives
\begin{eqnarray}
\nonumber \psi(\vec\theta^{k}) = \sum_{n_1=0}^{n_{\rm max}}
\sum_{n_2=0}^{n_{\rm max} - n_1}
\frac{\partial^{n_1}}{\partial\theta_1^{n_1}}
\frac{\partial^{n_2}}{\partial\theta_2^{n_2}} \psi(\vec\theta^{0})
\times\\ \frac{\round{\theta^{k}_1-\theta_1^0}^{n_1}
\round{\theta^{k}_2-\theta_2^0}^{n_2}}{n_1! n_2!}\; ,
\end{eqnarray}
where the maximum  derivative order $n_{\rm max}$ is determined 
by the measurements available to us. For example, if
our measurements were to include reduced flexion (e.g. \citealp{GL/Bac++06,
schneider08}), then we would set $n_{\rm max} = 3$ since we would
need third order derivatives of $\psi$.

This gives us a set of $N$ linear equations with $M=\frac{(n_{\rm
    max}+1)(n_{\rm max}+2)}{2}$ unknowns (which are the $\psi$ and its
derivatives at $\vec\theta^0$). We solve this system, requiring $N>M$
and using singular value decomposition (SVD) to invert the matrix.
This method is very robust in finding a solution for any of the
neighbour sub-grid shapes we investigated.  Note, however, that
choosing $N=M$ provides rather noisy estimates, especially if $n_{\rm
  max}$ is large, and occasionally the resulting set of equations
cannot be inverted. In practice, this situation is easily avoided by
setting $N>M$.  We repeat the SVD procedure for all points
$\vec\theta^0$ where we either have weak or strong lensing galaxies,
since this is where we need to be able to predict deflection angles
and reduced shear $g = \gamma / (1 - \kappa)$. The resulting solution
for the spatial derivatives of $\psi$ is used to
evaluate $\kappa$, $\gamma$, and $\vec \alpha$. We do not
measure higher order distortions such as reduced flexion from the
current data, although the method described above can clearly be
applied to those quantities as well. Note that the SVD procedure is only run
once, to derive the numerical formulae for the predicted observables
as a function of the gridded potential. All subsequent iterations
where the potential values are varied make use of these formulae.

The independent strong and weak lensing data are then combined by
multiplying their likelihoods together.  We compute the $\chi^2$
difference between the data (the positions of the multiple images, and
the ellipticities of the weakly lensed galaxies) and their model
predictions.  Since the weak lensing data is noisy and we choose to
work on a many-parameter lens potential pixel grid, regularisation
needs to be employed. This process ensures that unphysical
pixel-to-pixel variations in surface mass density are suppressed, and
corresponds to asserting a smoothness prior on the reconstruction.
In practice we compare the current surface
mass density and shear map with those of an initial model, 
and penalize strong deviations in their mean square difference by adding this
term to the data $\chi^2$
(see \citealp{bradac04a} for details). 
The regularisation is chosen (and tested to be) such as to give negligible
bias in the resulting aperture mass estimates.

The final global merit function (including data $\chi^2$ and
regularisation term) is equal (up to a normalising constant) to the
negative logarithm of the posterior probability density function (PDF)
for the model parameters $P$, the gridded potential values:
\begin{equation}
  -\log{P} = \frac{\chi^2_{\rm WL}(\psi)}{2} + \frac{\chi^2_{\rm SL}(\psi)}{2} + 
             \eta R(\psi) + {\rm constant} 
\end{equation}
\citep[see][for details]{bradac04a}.
%
The posterior peak values of the potential
$\psi_k$ are found by solving the non-linear equation $\partial
\log{P} / \partial \psi_k =0$.  We linearize this set of equations and
reach a solution in an iterative fashion (keeping the non-linear terms
fixed at each iteration step). This requires an initial guess for the
gravitational potential; the systematic effects arising from various
choices of this initial model were discussed in \citet{bradac06}.

We do not evaluate the covariance matrix of the multivariate Gaussian,
arguing that the mass reconstruction is actually limited by systematic
effects rather than statistical uncertainty.  One could numerically
evaluate the second derivatives of $\log{P}$ at the peak, or perhaps
estimate the errors on the final reconstruction using MCMC sampling
\citep{marshall06,jullo09}.  However with $N_{\rm grid} \gtrsim 3000$
this is computationally impractical at the moment and is beyond the
scope of this paper.

Instead we focus on systematic uncertainties in the magnification in
the inner ($3\arcmin \times 3\arcmin$) region around the main cluster,
where we later search for high-redshift galaxies, and which are
dominated by unmodeled substructure. We investigate the errors on
magnification in Sect.~\ref{sec:dropouts} by adding small scale mass
clumps  (as well as varying the smooth mass profile) to
the reconstruction and investigating their effect. This is adequate
for this work: the accuracy with which our lens potential model
predicts the positions of multiple images gives us an indication of
the range of allowable perturbations.

The choice of particular grid geometry, the regularisation parameter,
and the hyper-parameters that set the relative weighting between the
contributions to $\chi^2$ all become critical when weak lensing data
on large scales ($\gtrsim 1\mbox{Mpc}$) are included and we need a
full-field mass reconstruction. This is not the case in this work, as
we are only interested in the magnification of the inner region. We
leave this investigation to future work, where a complete mass
reconstruction of the cluster will be presented.


\section{Observations and data reduction for the Bullet Cluster data}
\label{sec:data}

ACS/WFC imaging of the Bullet Cluster was carried out in Cycle 13
(proposal 10200, PI Jones) on 2004 October 21 and Cycle 15 (proposal
10863, PI Gonzalez) on 2006 October 12-13. The data consists of two
pointings centered on the main cluster and the subcluster with a small
overlap between them. The main cluster has been imaged in F606W
(hereafter $V$, 2340s), F775W ($i$, 10150s), and F850LP ($z$, 12700s), while
the subcluster has been imaged in F435W(2420s), F606W (2340s), and
F814W (7280s). For the high redshift galaxy survey described below
we primarily use the main cluster's $Viz$ data.
The rest of the data was used in addition when generating the 
weak and strong lensing data catalogs as described in \citet{bradac06}.

We use the {\tt Multidrizzle} \citep{multidrizzle} routine to align
and combine the images. To register the images with the astrometric
accuracy needed for lensing analysis, we determine the offsets among
the images by extracting high $S/N$ objects in the individual,
distortion corrected exposures. We use {\tt SExtractor}
\citep{sextractor} and the IRAF routine {\tt geomap} to identify the
objects and calculate the residual shifts and rotation of individual
exposures, which were then fed back into {\tt Multidrizzle}. We use
``square'' as the final drizzling kernel and an output pixel scale of
$0.03\mbox{ arcsec}$; this is smaller than the original pixel scale of
the ACS CCD, allowing us to exploit the dithering of the observations
and improve the sampling of the PSF.  The limiting magnitudes were
  estimated using $0.2''\times0.2''$ appertures. The $5-\sigma$ limiting
  magnitudes are 28.2 for $V$, 27.6 for $i$ and 27.4 for $z$-band
  data.


\section{Improved mass reconstruction for the Bullet Cluster}
\label{sec:results}

\begin{figure}[ht]
\begin{center}
\begin{minipage}{0.5\textwidth}
\begin{tabular}{ccc}
\includegraphics[width=0.3\textwidth]{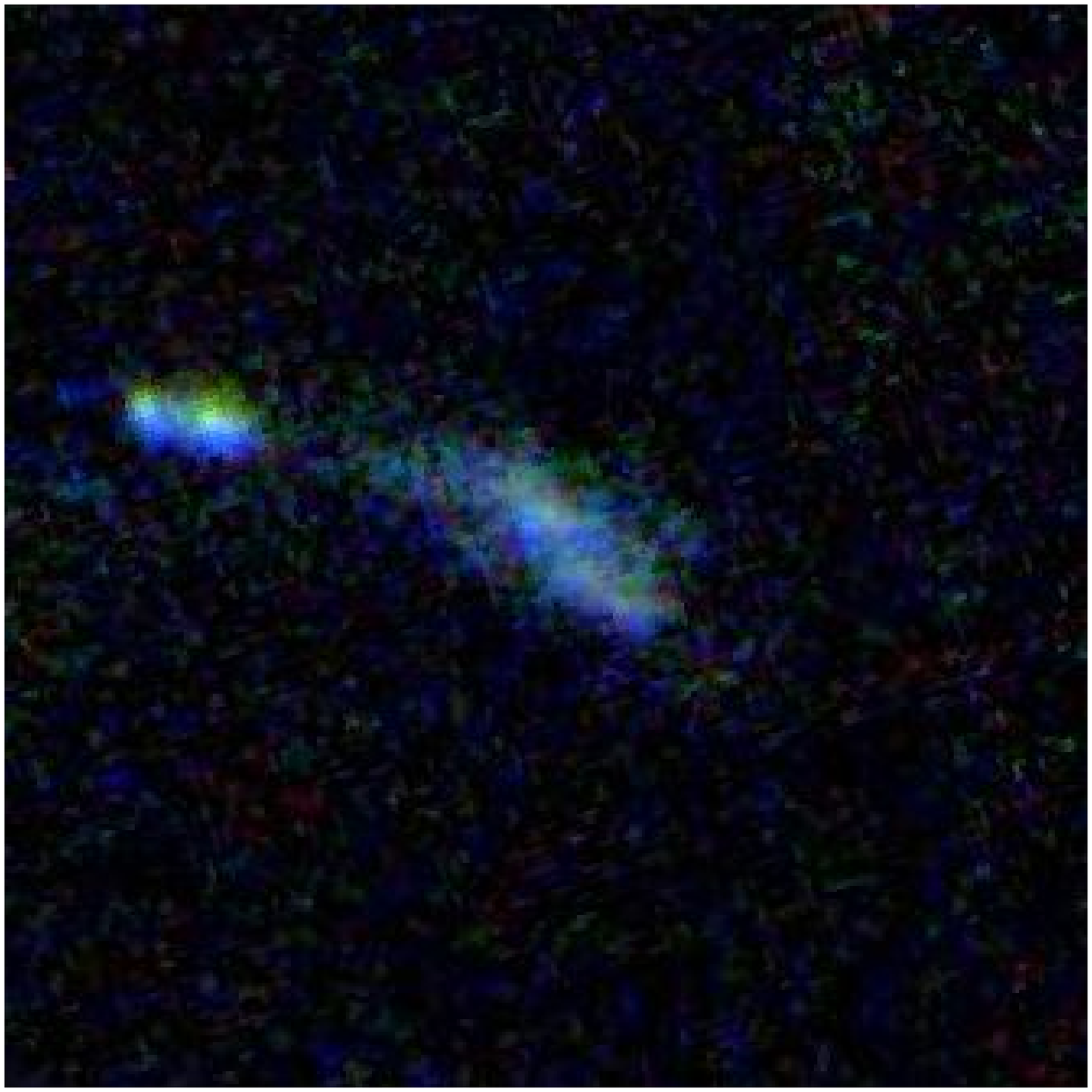} &
\includegraphics[width=0.3\textwidth]{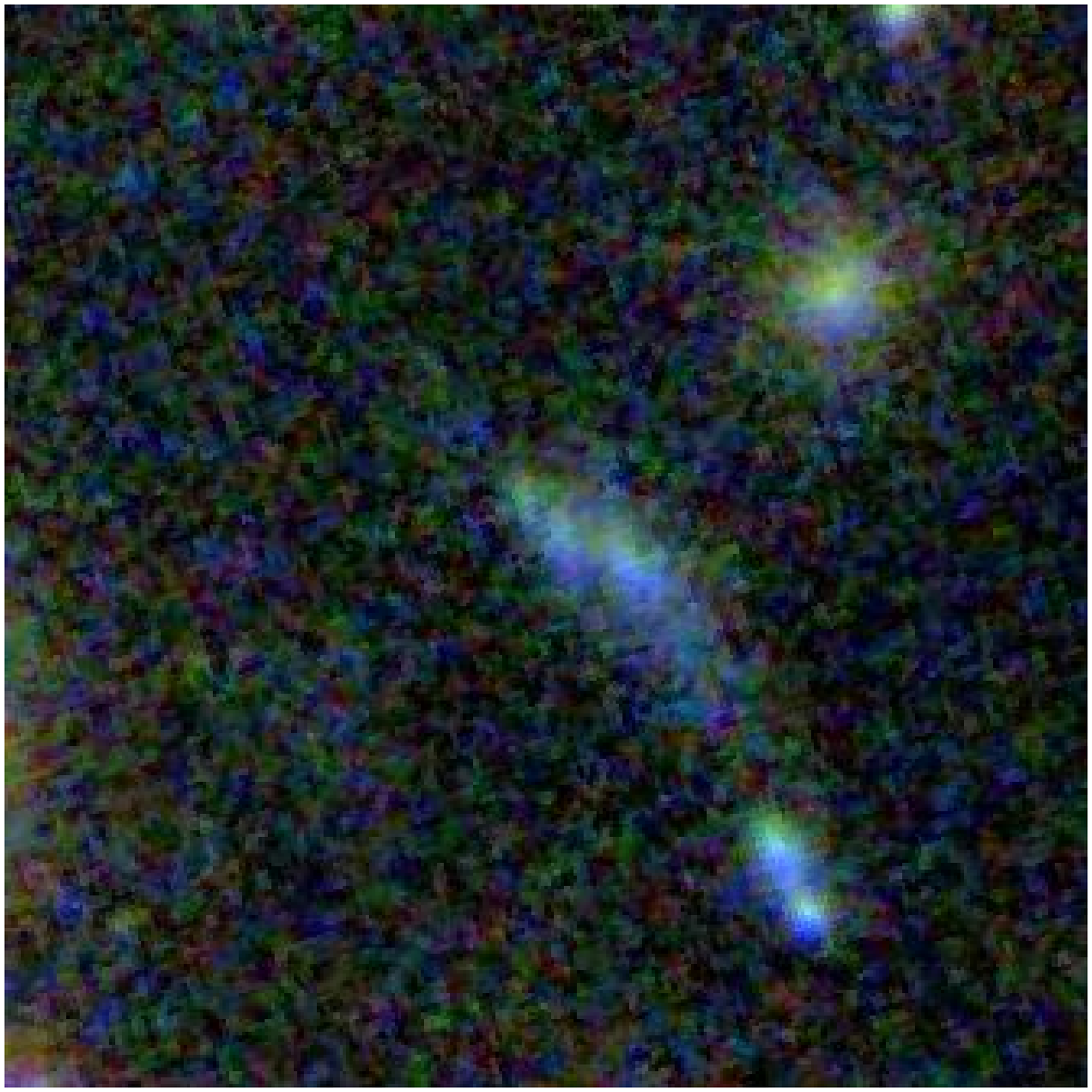} &
\includegraphics[width=0.3\textwidth]{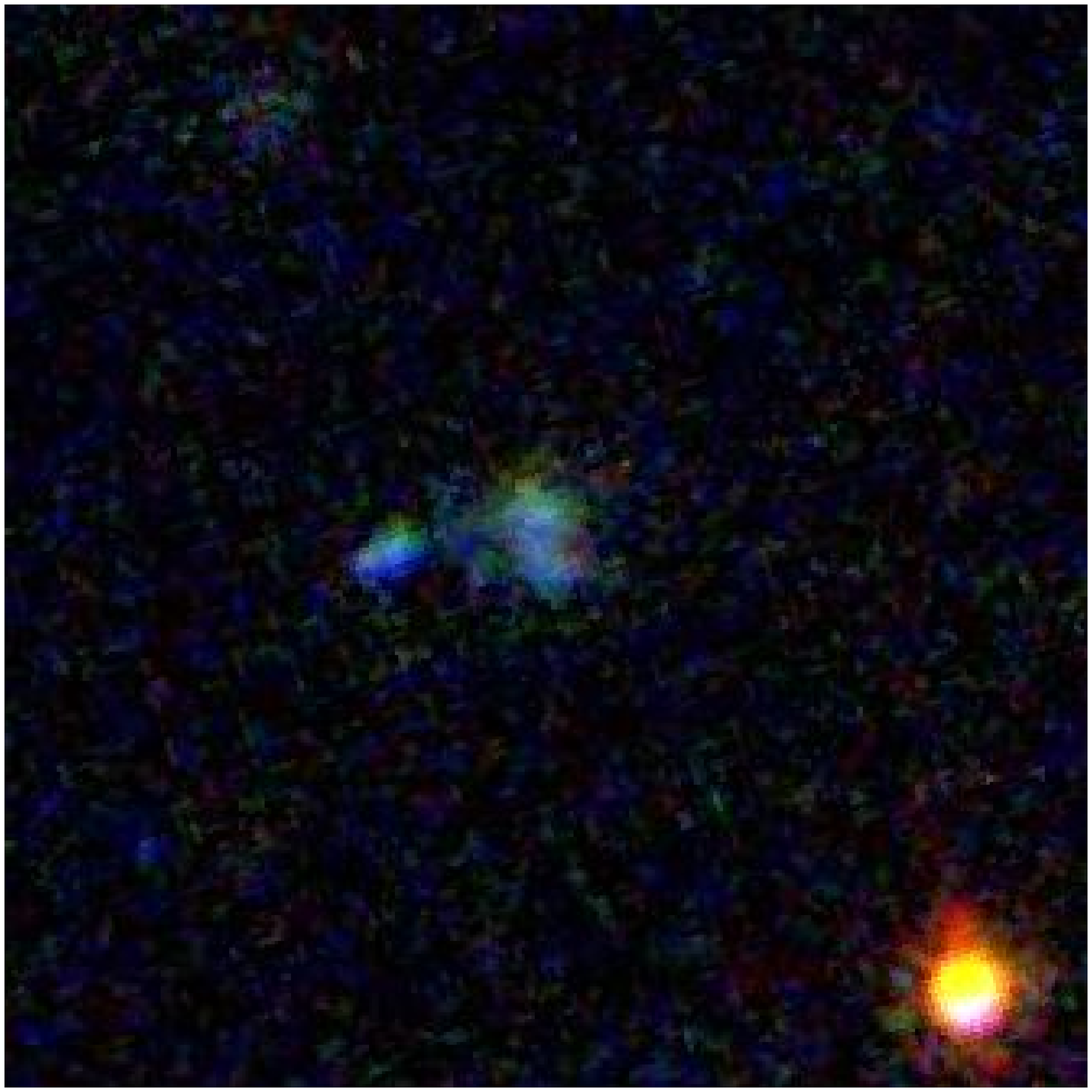} \\
\multicolumn{3}{c}{%
\begin{tabular}{cc}
\includegraphics[width=0.3\textwidth]{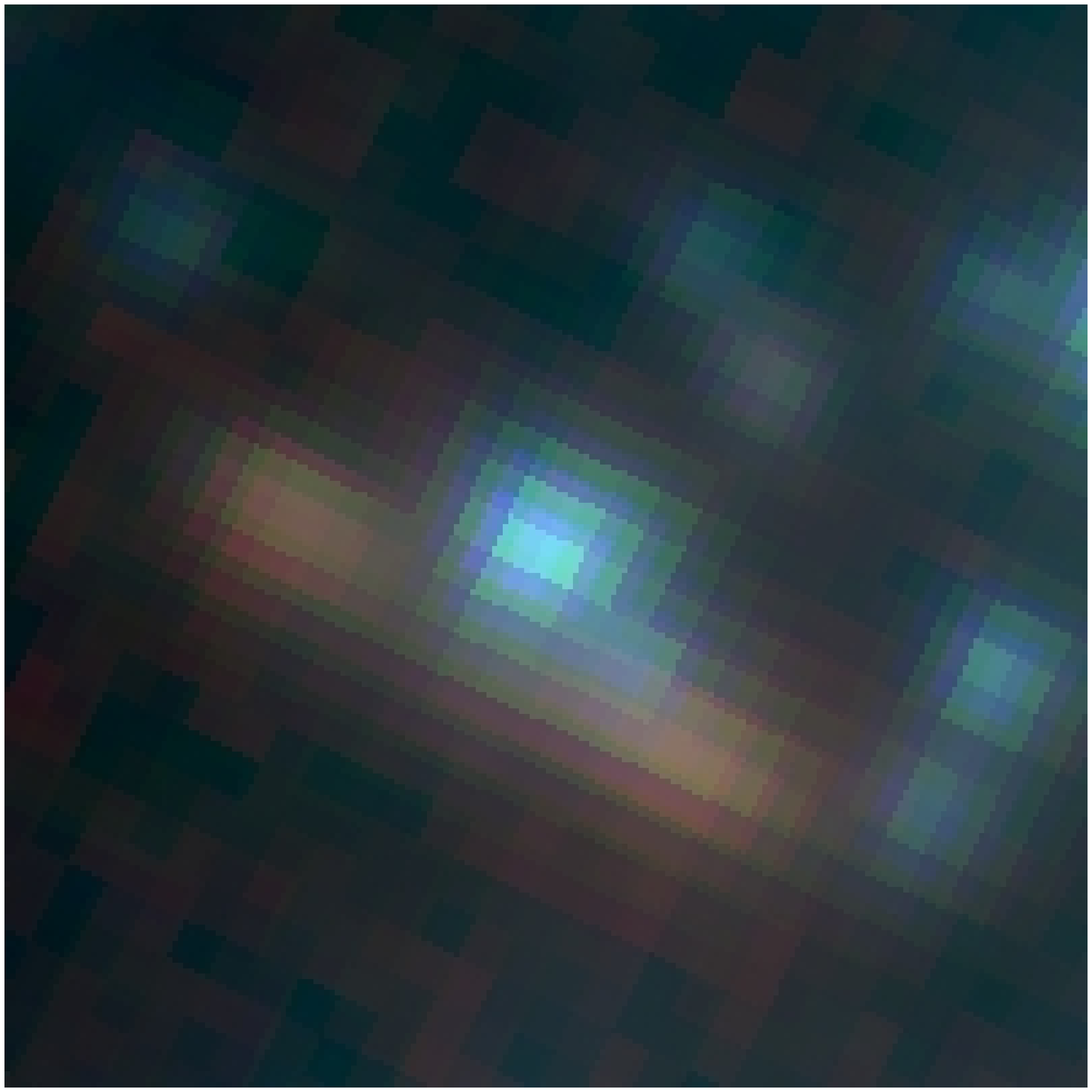} &
\includegraphics[width=0.3\textwidth]{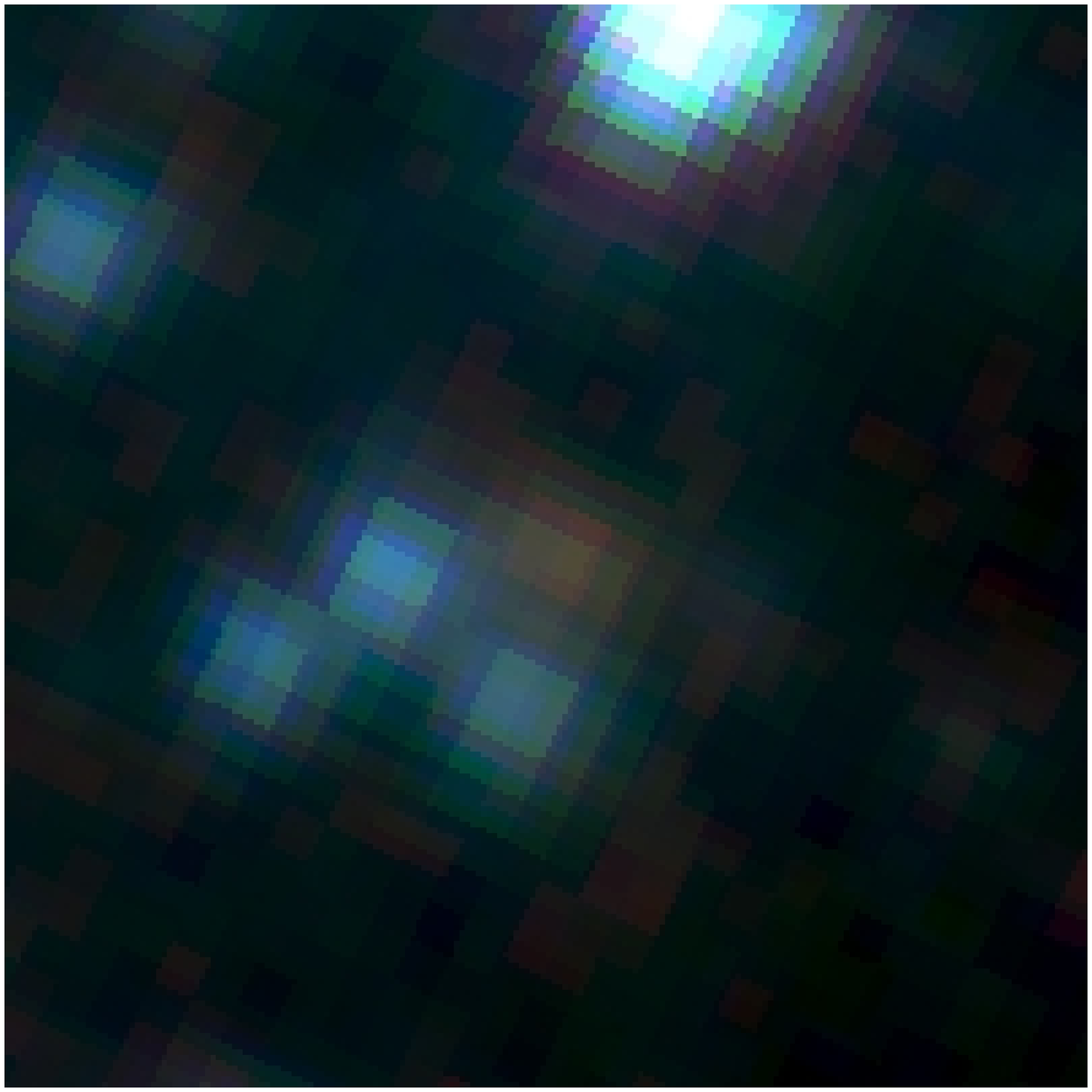} \\
\end{tabular}}\\
\multicolumn{3}{c}{%
\begin{tabular}{cc}
\includegraphics[width=0.3\textwidth]{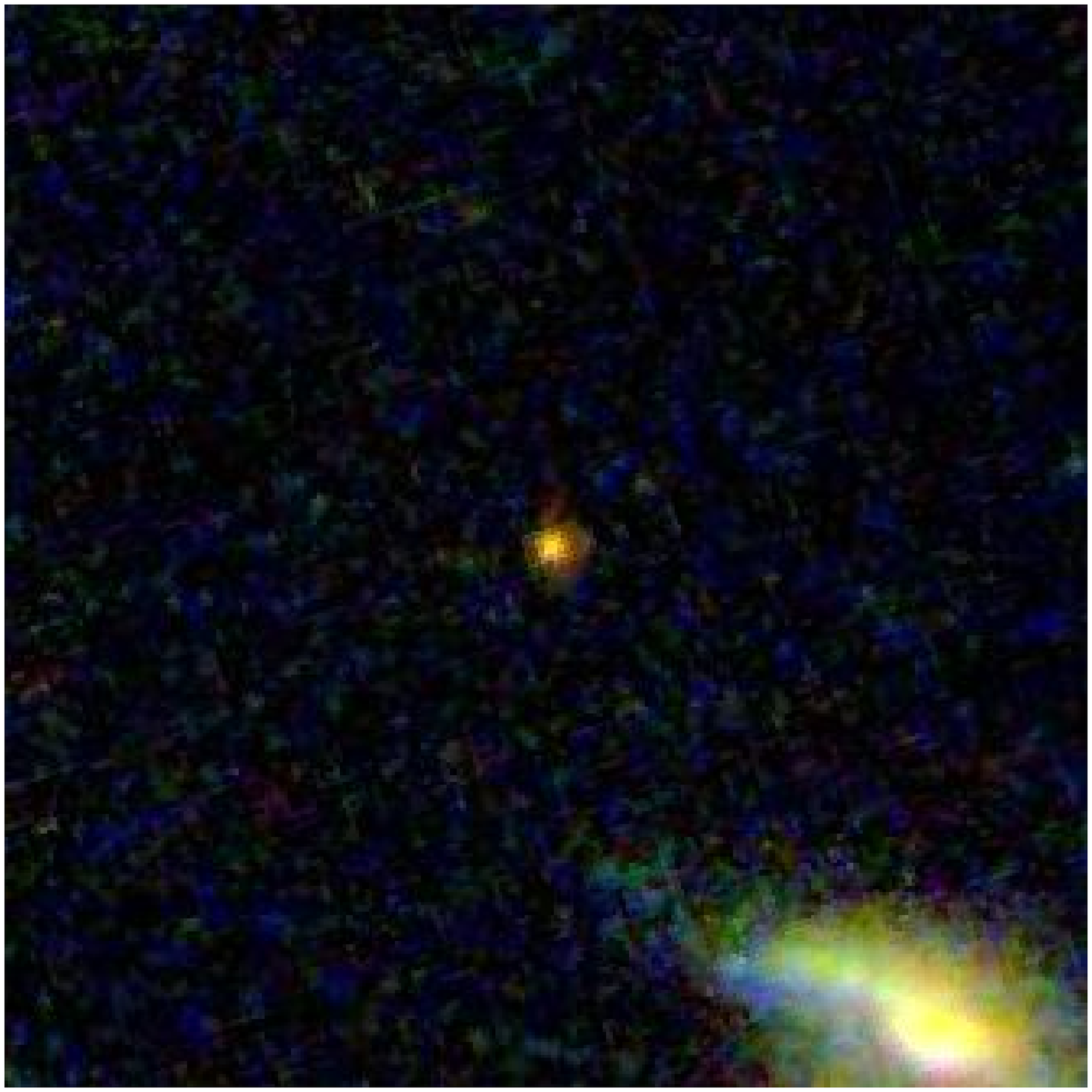} &
\includegraphics[width=0.3\textwidth]{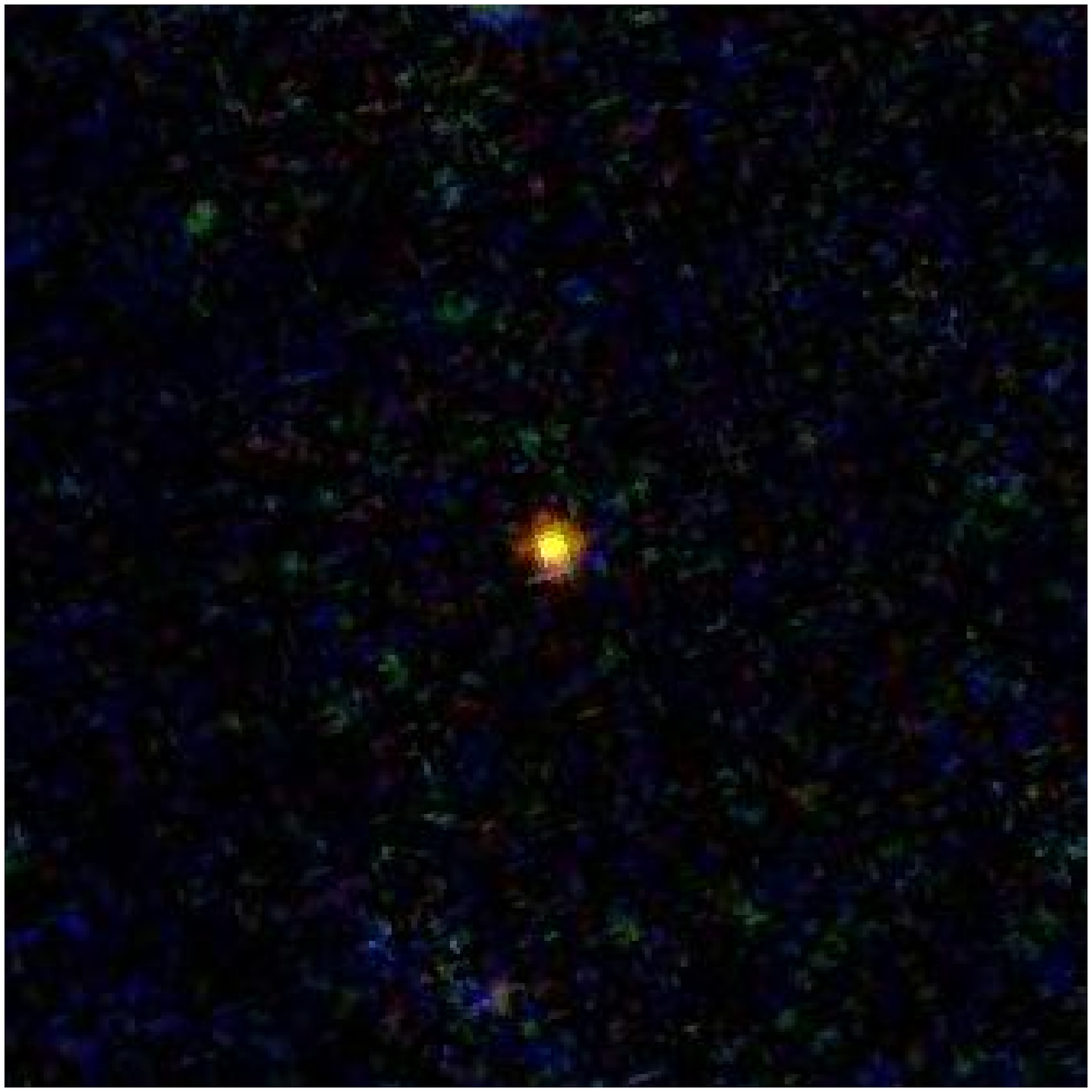} \\
\end{tabular}}
\end{tabular}
\end{minipage}
\end{center}
\caption{Color composites of the three multiple image systems: F (top)
  and the two of K (middle) and L (bottom). $Viz$ color composite is
  used for systems F and L. System K is not detected in any of the
  optical bands; instead, we show Spitzer/IRAC
  $3.6\mu\mbox{m}$-$4.5\mu\mbox{m}$-$8\mu\mbox{m}$ images (see also
  \protect\citet{gonzalez09}. North is up and East is left in each
  panel.  The cutouts are $10\arcsec\times 10\arcsec$.}
\label{fig:arcs}
\end{figure}

For the strong and weak lensing mass reconstruction, we use the deep
\HST/ACS images described above and the data used in \citet{bradac06}
and \citet{clowe06}. We use the catalog of strongly lensed images from
\citet{bradac06}, with the addition of two new multiple imaged systems
and one additional image (see Table~\ref{tab:arcs}). The weak lensing
catalogs are taken from \citet{clowe06} and are obtained from a
combination of ACS and ground-based data and extends to a large FOV;
we use the inner $9\arcmin\times9\arcmin$ here.  The positions and
redshifts of the strongly-lensed images are given in
Table~\ref{tab:arcs} (see also Figure~\ref{fig:arcs}).  The first new
multiply imaged system (Table~\ref{tab:arcs}, K) is the IRAC-selected
system reported in \citet{bradac06} for which \citet{gonzalez09}
estimated a photometric redshift of $z=2.7 \pm 0.2$ and also found an
additional image.  The second new system (Table~\ref{tab:arcs}, L) is
a multiply imaged pair at $z\sim 5$ discovered in our search for
$z\sim 5-6$ galaxies and described in Section~\ref{sec:dropouts}. We
also include a new counter image for system F, which we found in close
proximity to the position predicted by our lens model.

\begin{figure}[ht]
\begin{center}
\includegraphics[width=0.45\textwidth]{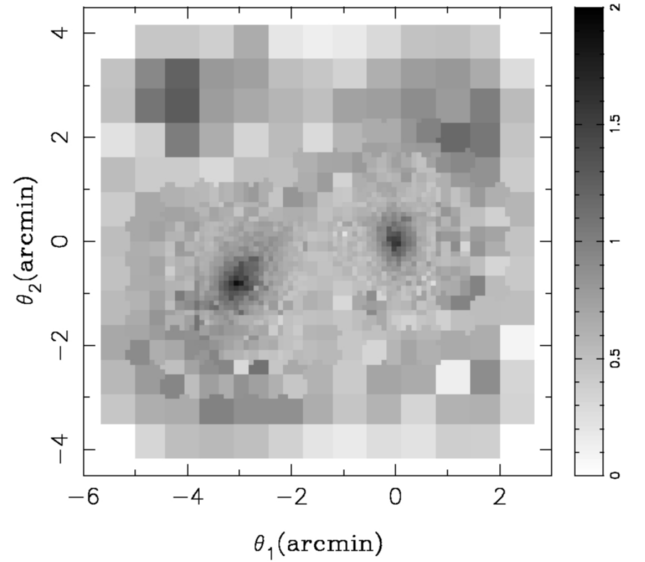}
\end{center}
\caption{The reconstructed surface mass density $\kappa$ of the
  cluster for a fiducial source at infinite redshift, $z_{\rm s} \to
  \infty$.}
\label{fig:kappamap}
\end{figure}

The reconstruction is set up on an initial grid of $15\times15$ pixels
for a field of $9\arcmin \times 9\arcmin$.  We then adapt the grid as
follows. We have the highest S/N ratio in areas close to the observed
multiple images; in addition, the S/N ratio declines as we go towards
the outskirts of the cluster.  We therefore refine the circular
regions of $1\arcmin$ diameter around the centers of the main and the
subcluster by a factor of 8 (i.e. each grid is split into $8\times 8$
grid points), by a factor of 4 in an annuli between $1\arcmin$ and
$1.5\arcmin$, and by a factor of 2 in an annuli between $1.5\arcmin$
and $2\arcmin$. We also refine cells surrounding $0.5\arcmin$ around
each multiply imaged system by 16. The resulting grid structure is visible
in Fig.~\ref{fig:kappamap}: the smallest cells have sizes of
$2.25\arcsec$, corresponding to $\sim 10\mbox{ kpc}$ and to the Einstein diameter of massive ellipticals.

\begin{figure*}[ht]
\begin{center}
\includegraphics[width=0.8\textwidth]{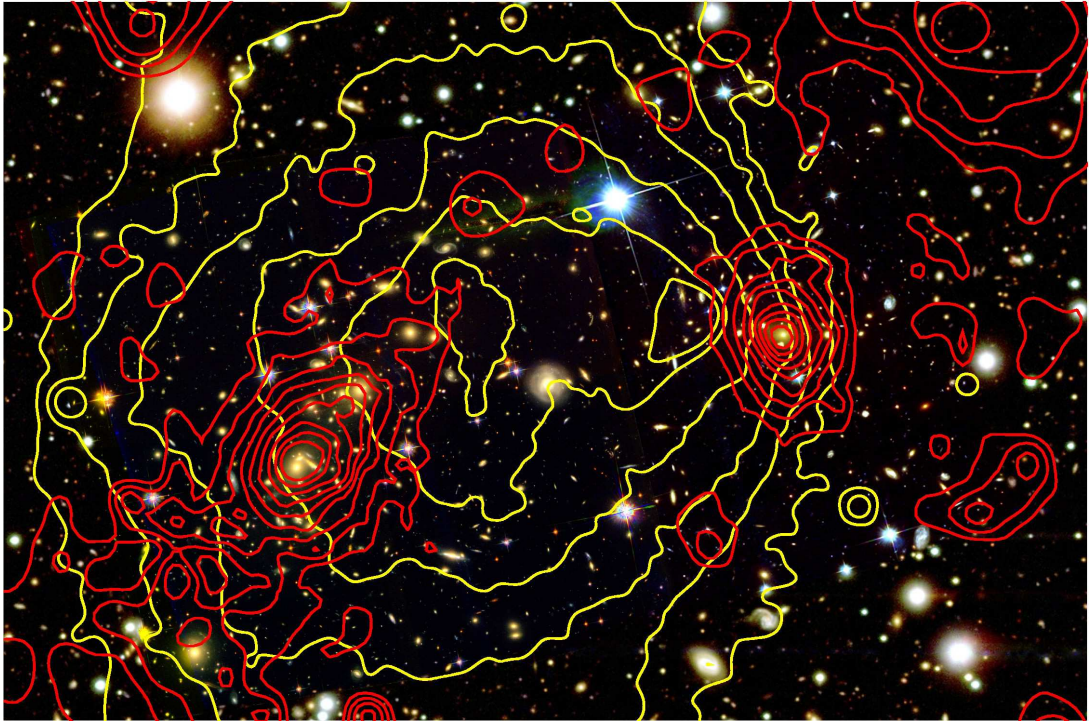}
\end{center}
\caption{The $Viz$ (main cluster) and F435W-$V$-F814W (sub cluster) color
composite of the \protect \bullet. The ACS images are inset on the
colour image from NASA Press release 06-297.  Overlaid in {\it red contours}
is the surface mass density $\kappa$ from the combined weak and strong
lensing mass reconstruction.  The contour levels are linearly spaced
with $\Delta\kappa = 0.1$, starting at $\kappa=0.7$, for a fiducial
source at a redshift of $z_{\rm s} \to \infty$. The X-ray brightness
contours from the 500 ks Chandra ACIS-I observations
\citep{markevitch06} are overlaid in {\it yellow}. North is up and
East is left, the field is $7.4^{\prime}\times 5.3^{\prime}$, which
corresponds to $2000 \times 1400 \mbox{ kpc}^2$ at the redshift of the
cluster. The color composite was created following the algorithm from
\citet{lupton04} and using the image from NASA Press release 06-297.}
\label{fig:swunited}
\end{figure*}

Only for system A has a spectroscopic redshift been obtained
\citep{mehlert01}.  Hence, we first perform the initial reconstruction
using redshifts given in \citet{bradac06}, which are a combination of
photometric redshifts and lensing predictions from a smooth,
simply-parametrized model. Then, we project all the images for each
source to the source plane, take the average source position and
calculate the rms difference between the measured image positions and
those predicted by the model for the average source. We then vary the
source redshift (while keeping the mass reconstruction fixed) around
its predicted value given in \citet{bradac06}. The redshift with the
smallest rms for each of the sources is given in
Table~\ref{tab:arcs}.  The final surface mass density maps is
  given in Figure~\ref{fig:swunited}, together with the X-ray
  brightness contours. The average rms in predicted image positions
per image (and source) is $1.4\arcsec$. The reconstruction has
improved significantly with respect to our earlier map (in
\citealp{bradac06} the average rms was $4\arcsec$). The rms residuals
depend upon the final adopted grid size. Whereas the final grid size
could be further decreased, we chose not to do so, as the lack of
redshift information limits our ability to reconstruct these small
scales. We can achieve similar fits by adding small scale substructure
close to the images, or changing their redshifts.  Hence, in order to
improve the mass reconstruction even further, we will need
spectroscopic redshifts for more than one system, which are not
presently available.

The statistical uncertainties in the redshifts of the multiply imaged
sources can be estimated while keeping the mass reconstruction
fixed. However, these are small and do not represent the true
uncertainties. We would need to simultaneously vary the lens model and
the redshifts to obtain true errors; this, however, is extremely
computationally intensive. At present, however we mimic redshift
uncertainties by adding small scale mass perturbations to the model in
Section~\ref{sec:dropouts}. As the two are degenerate, the
perturbations that produce the shifts in the predicted positions of
the multiple images larger than the rms residuals quoted above will give us
conservative errors on the model.

\begin{deluxetable}{rrrr}
\tablecolumns{4}
\tablecaption{The properties of the multiply-imaged systems used in
  this work.}
\tablehead{ \colhead{} &  \colhead{Ra} & \colhead{Dec} & \colhead{$z_{\rm pred} (z_{\rm pred06})$}}
\tablewidth{0pc}
\startdata
&104.65852&$-$55.950585&  \\
K&104.65447&$-$55.951904&2.8 (N/A)\\
 &104.63917&$-$55.958060&\\
\cline{1-4}
&104.65024&$-$55.961577&\\
\raisebox{0.5ex}{L}&104.64327&$-$55.963856&\raisebox{0.5ex}{5.7 (N/A)}\\
\cline{1-4}
&104.64694 &$-$55.958465 & \\
F\tablenotemark{1} &104.65209 &$-$55.956241& 0.9 (0.8)\\
&104.66497 &$-$55.951324 & \\
\hline\hline
& 104.63316 & -55.941395 & \\
\raisebox{0.5ex}{A} & 104.62988 & -55.943798 & \raisebox{0.5ex}{3.24 (3.24) \tablenotemark{2}} \\
\cline{1-4}
& 104.62954 & -55.941844 & \\
\raisebox{0.5ex}{B} & 104.63042 & -55.941474 & \raisebox{0.5ex}{3.9 (4.8)} \\
\cline{1-4}
& 104.63775 & -55.941851 & \\
\raisebox{0.5ex}{C} & 104.63338 & -55.945324 &  \raisebox{0.5ex}{2.3 (2.1)} \\
\cline{1-4}
& 104.64709 & -55.943575 & \\
\raisebox{0.5ex}{D} & 104.63528 & -55.951836 &  \raisebox{0.5ex}{1.5 (1.4)} \\
\cline{1-4}
& 104.64008  &  -55.950620 & \\
\raisebox{0.5ex}{E} & 104.64232  &   -55.948784&  \raisebox{0.5ex}{0.9 (1.0)} \\
\cline{1-4}
& 104.56568  & -55.939832 & \\
G & 104.56402& -55.942113 &1.1 (1.3)\\
& 104.56417 & -55.944131  & \\
\cline{1-4}
 & 104.56293 & -55.939764 & \\
H & 104.56133 & -55.942430 & 2.1 (1.9)\\
 & 104.56189 & -55.947724 & \\
\cline{1-4}
&104.56186& -55.946114  & \\
I &104.56052 &-55.942930 & 2.3 (2.1)\\
&104.56141 &-55.944264 & \\
\cline{1-4}
& 104.56909 & -55.946016  & \\
\raisebox{0.5ex}{J} & 104.57025 & -55.944050 & \raisebox{0.5ex}{1.3 (1.7)}
\enddata
\tablenotetext{1}{First two images are from \citet{bradac06}, the last was predicted in its close proximity by the lens model.}
\tablenotetext{2}{ Only image A has a measured spectroscopic redshift \protect \protect \citep{mehlert01}. Others are the photometric redshifts refined by the gravitational lensing model. Systems from here on were used already in \protect \citet{bradac06}, the redshifts used there are listed in parenthesis.}
\label{tab:arcs}
\end{deluxetable}


\section{Bullet Cluster as a Cosmic Telescope: Lyman Break Galaxies
at $z\sim 5-6$}
\label{sec:dropouts}

In this section we present a search for $V$ and $i$-band dropout
galaxies behind the Bullet Cluster.


\subsection{Dropout selection and photometry}

Galaxies were detected in the \HST/ACS $Viz$ imaging data described in
Section~\ref{sec:data}; we followed a procedure almost identical to
that of \citet{stark09} and \citet{coe06}.  Faint objects were
detected from a combined $Viz$ image, and colors measured in apertures
of $0.6^{\arcsec}$ diameter. The foreground galactic extinction was
corrected following \citet{ned1} and \citet{ned2}. We also applied an
aperture correction for the F850LP image of $-0.05$ following
\citet{coe06}. The total magnitudes in each band were computed using a
combination of aperture color and {\tt SExtractor} parameter MAG\_AUTO
from the reddest filter F850LP.  We then selected filter dropouts with
the same criteria as used in \citet{stark09} (and also in \citet{beckwith06}). For
$V$-dropouts we require all of the following:
\begin{eqnarray}
        (V - i  > 1.47 + 0.89(i - z)) \;\;{\rm or}\;\;  (V - i >  2)& &  \\
      V- i > 1.2 & &\\ i - z < 1.3& & \\
      S/N(z) > 5& &
\end{eqnarray}
and since we do not have $B$-band data, we omit the last selection
criterion from \citet{stark09}. This does not significantly affect our
results, as there are very few possible objects that would drop out in
V and be detected in B (e.g. an AGN in a dusty galaxy). In addition,
the study of \citet{bouwens07}, which we also compare to, does not use
B-band non-detection either.  For $i$-dropouts we use:
\begin{eqnarray}
        i- z > 1.3 & &\\
            S/N(z) > 5  & &\\
         (S/N(V_{606}) < 2) \;\;  {\rm or}\;\; (V-z > 2.8) & &\:.
\end{eqnarray}

\begin{figure}[ht]
\begin{center}
\includegraphics[width=0.5\textwidth]{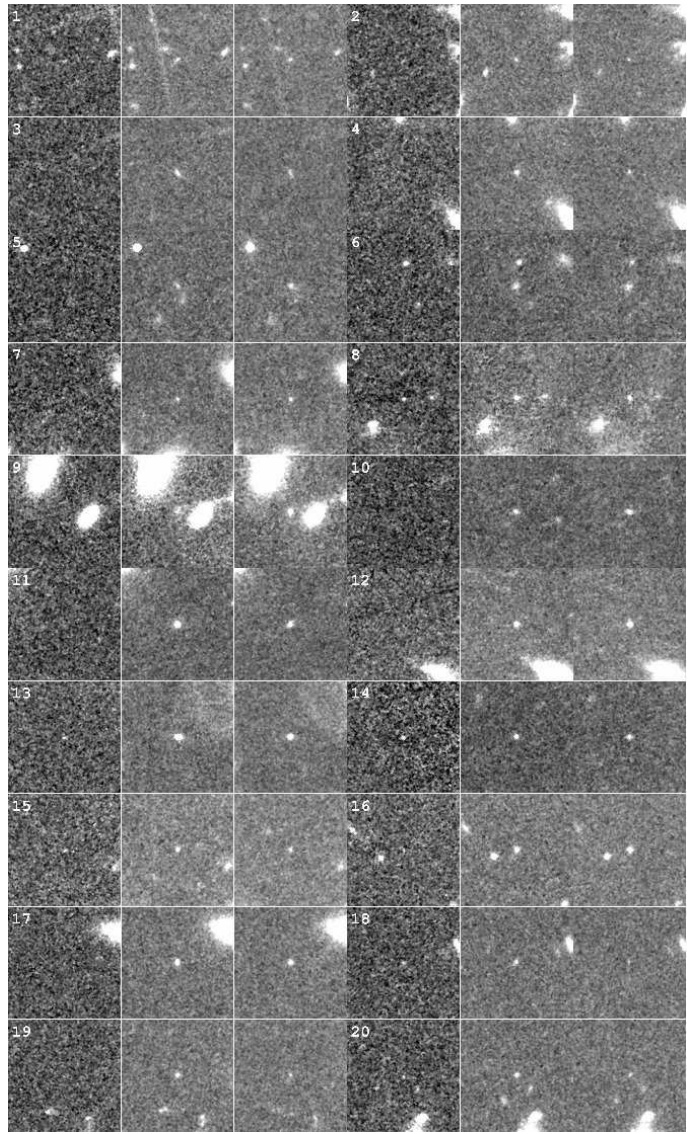}
\end{center}
\caption{Cutouts of $V$-band dropouts. The cutouts are shown for
  filters $Viz$ (in two columns), labels correspond to labels in
  Table~\ref{tab:dropouts}. The sizes are $10\arcsec\times
  10\arcsec$.}
\label{fig:vdropsim}
\end{figure}

\begin{figure}[ht]
\begin{center}
\includegraphics[width=0.25\textwidth]{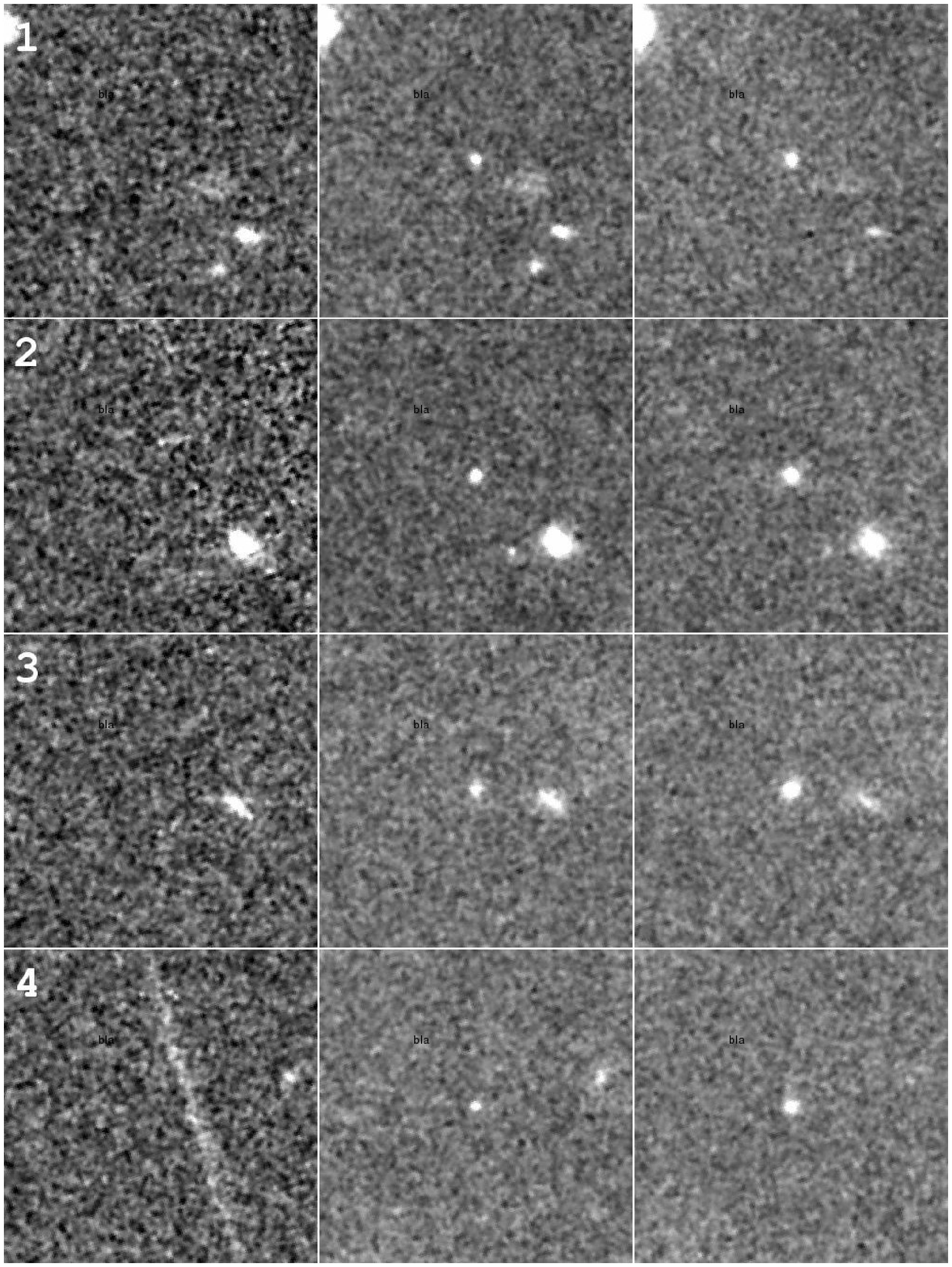}
\end{center}
\caption{Same as Fig.~\ref{fig:vdropsim}, except for $i$-band dropouts.}
\label{fig:idropsim}
\end{figure}

In addition, we rejected stars using the {\tt IMCAT} software.  We
determined the significance and size of each object in each passband
by convolving the images with a series of Mexican hat filters and
determining the smoothing radius, $r_{\rm g}$, at which the filtered
objects achieved maximum significance.  This procedure, usually used
for weak lensing analyses (see e.g. \citealp{clowe06b}), gives a
better estimate of the size than {\tt SExtractor}.  We then identified
stars from the $r_{\rm g}$ vs.\ magnitude diagram and rejected the
bright stars ($\magz < 22$). Finally, we visually examined each object
from the catalog in the images, removing any additional faint stars,
artifacts (at the detector edges) from the sample.  The visual
  inspection is successful in removing stars with ($\magz < 26$), at
  magnitudes 26-27, however, the fraction of stellar contaminants is
  negligible \citep{bouwens06}. Our final sample comprises 20 $V$ and
  4 $i$-band dropouts, their positions are given in
  Table~\ref{tab:dropouts} and the cutout images are shown in
  Figures~\ref{fig:vdropsim} and \ref{fig:idropsim}.

\subsection{Estimate of the completeness and ``lensed'' number counts}

To estimate the completeness of the sample we ran simulations using
IRAF task {\tt artdata}. We generated 1000 objects each at the average
$z$-band magnitude $\magz$ of $\{24,25,26,27,28\}$, with colors
corresponding to a template spectrum from \citet{bruzual93} for a low
metallicity ($Z=0.4\:Z_{\odot}$) starburst galaxy with an age of
$125$~Myr and no dust. We assigned random redshifts to the galaxies
(uniformly distributed in $4.0<z<5.9$ for $V$, and $5.0<z<7.6$ for
$i$-band dropouts), and then follow \citet{madau96} in applying a
corresponding IGM absorption correction. The galaxies were added to
the $Viz$ images of the cluster and detected following the same
procedure.  The completeness levels are $>95\%$ for $\magz = 24$,
  90\% for $\magz = 25$, 50\% for $\magz = 26$, and 5\% at $\magz =
  27$. To estimate the
  completeness for each individual object having $\magz < 27$, a
  fourth order polynomial as a function of $\magz$ was fitted through
  above estimated completeness levels. The objects were then later
  added to the appropriate bins, assuming these completeness
  factors. We exclude any objects having (lensed, measured) magnitudes
  $\magz \ge 27$ when calculating numbercounts of objects as a function
  of magnitude (see Fig.~\ref{fig:numcount}). The size of the field is
$10.24\mbox{ arcmin}^2$.

We can now calculate the surface densities for our two dropout samples
and compare them to values from the literature (we choose
\citealp{stark09, bouwens07, beckwith06} as they use the same
filter-set and a comparison is straightforward).  We bin our data
  in two bins of $\magz = [25-26]$ and $[26-27]$ for $V$ and one bin
  $\magz = [25-26.5]$ for $i$-dropouts. For the comparison with the
  literature the bins are then divided by 2 and 3 respectively to
  obtain counts in $0.5$-mag bins.  Our results are presented as grey
open circles with dashed errorbars in Fig.~\ref{fig:numcount} (the
determination of the ``unlensed'', true number counts ---shown as
black solid circles in Fig.~\ref{fig:numcount}--- will be described in
Section~\ref{sec:numcmag}).  The errors include both Poisson errors
and sample variance (also commonly referred to as cosmic variance).
To calculate the sample variance we use the prescription from
\citet{somerville04} and the correlation function from
\citet{nagamine07}. The sample variance for a single ACS field at
$z\sim 6$ is $50\%$, which is also in agreement with sample variance
calculator by \citet{trenti08}.  We immediately note that the lensing
effect has increased the number of detected dropouts compared to blank
surveys with similar or better exposure times to ours.  Even though
gravitational lensing effectively makes the actual observed solid
angle smaller (due to magnification), the effective slope of the
luminosity function $\Phi$ is steep at these magnitudes ($-\rm d(\log
\Phi) / d(\log L) \gtrsim 1$), hence we increase the expected
number density of sources by going deeper.


\subsection{Number counts including magnification correction}
\label{sec:numcmag}

To evaluate the true (unlensed) number counts, we use the mass
reconstruction described in Section~\ref{sec:results} to estimate the
effective change in survey area.  We use the deflection angle map for
a redshift $z=5$ source for the $V$-dropouts, and for a $z=6$ source
for the $i$-dropouts. We project the observed field from the lens
plane to the source plane pixel by pixel and measure the corresponding
change in solid angle by simple numerical integration.  This resulted
in a fractional area decrease of $0.25 \pm 0.02$ for $z=5$ and $z=6$
case. The total shrinking of the field changes little with redshift,
as the lensing strength (or angular diameter distance ratio between
lens and source, and observer and source) at these redshifts for a
lens at $z\sim 0.3$ is nearly constant (see e.g
\citealp{bartelmann00}). The uncertainty was estimated by evaluating
the area change using the magnification error map (described
below). Neither a source redshift change (supported by the calculated
solid angle change for sources at $z=5$ and 6) nor the magnification
errors have a significant effect on the survey area shrinkage.  The
position of the critical curves changes by $\sim 5\mbox{ arcsec}$
between $z=5$ and $z\to\infty$, which is larger than the expected
accuracy; even so, this has an almost negligible effect (within the
errors quoted above) on the total solid angle change of the field.

Next we need to evaluate what the apparent magnitudes of the sources 
would have been in the
absence of lensing. For this, we evaluate individual magnifications at
the positions of these sources using our lens model. 
The errors on each magnification come from
the errors on the mass reconstruction, 
and typically increase as the image approaches
the critical curve. These errors are dominated by small-scale substructure not
accounted for in the lens modeling, and by erroneous redshifts
of multiple images used for the reconstruction.  They are less affected by the 
typical uncertainties in cluster mass profile as discussed below.

\subsection{Errors on magnification}
\label{sec:errmag}

To estimate the errors we used (for convenience) the approximate
parametrized model for the Bullet Cluster (consisting of one PIEMD
(pseudo-isothermal elliptical mass density, see
e.g. \citealp{limousin05}) component for each of the main and
subcluster, as well as 30 SIS (Singular Isothermal Spheres) placed
over the full mass-reconstruction field representing the cluster
galaxies (these are the confirmed cluster members in the inner
$5\arcmin \times 3 \arcmin$ field). These galaxies were taken to have
line-of-sight velocity dispersion $\sigma_{\rm memb}$ following
$\sigma_{\rm memb} \propto L^{1/4}$ and a fiducial galaxy with a
F606W-band magnitude $m_{\rm F606W}=18$ has $\sigma_{\rm memb, 18} =
250\kms$. We chose singular rather than non-singular isothermal
spheres, as they will create larger local magnification changes. In
addition, strong lensing studies of galaxy-scale lenses show that they
have steep mass profiles in the centre, and hence small core radii
(see e.g. \citealp{keeton03b}).

\begin{figure}[ht]
\begin{center}
\includegraphics[width=0.5\textwidth]{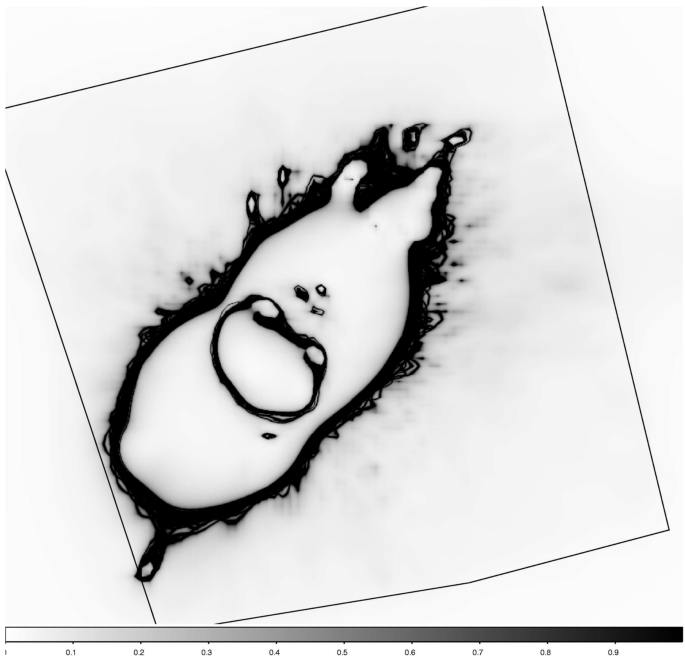}
\end{center}
\caption{The error on intrinsic (unlensed) source magnitude
  $\sigma_{\rm mag}$ (grey-shade) as a function of position. Most of
  the lens-plane area has errors $<0.1\mbox{ mag}$.  The field we
  study here is approximately centered on the main cluster and is
  given as a box.  The map assumes a source at $z\to\infty$, the
    map for $z=5$ or 6 would differ very little from the one presented
    here.}
\label{fig:errmap}
\end{figure}

The above model gives us a manageable smooth model for the cluster
mass distribution.  To model the errors on the magnification due to
substructure, we also randomly distributed 50 mass clumps with
position covering our $Viz$ images (where dropouts were searched for)
and magnitudes between $m_{\rm F606W}=22\mbox{ - }24$ (and the same
scaling of mass as used above). Again, we are trying to maximize the
magnification perturbations, hence using singular profiles is
conservative.  The mass fraction in these substructures was $\sim
8\%$; this is purposely high, since we want to be conservative in
estimating the magnification errors.  This model does not reproduce
the multiple image positions in detail (i.e. we did not search for the
best-fit model - the substructure fraction would have been lower and
systematic errors due to uncertainties in redshifts and the main lens
profile would be underestimated); however this is not needed for the
error estimation.  This additional substructure changes the critical
curve positions by $5-10\mbox{ arcsec}$, which is more than the
uncertainty with which multiple images are recovered (in this and in
other well-modeled clusters, see e.g. \citealp{jullo09}). We generated
100 realisations of cluster substructure in this way, and then
evaluated the magnification error as the standard deviation of the
mean magnification at each pixel. The map of standard deviation in the
change of magnitude $\sigma_{\rm mag} \simeq \sigma_{\mu}/\ave{\mu}$
as a function of position is given in Fig.~\ref{fig:errmap}. On
average the errors are small across the observable field, only in the
vicinity of the critical curve the errors become large. The errors at
the positions of the dropouts are given in Table~\ref{tab:dropouts}.

\begin{figure*}[ht]
\begin{minipage}{0.5\textwidth}
\begin{center}
\includegraphics[width=0.9\textwidth]{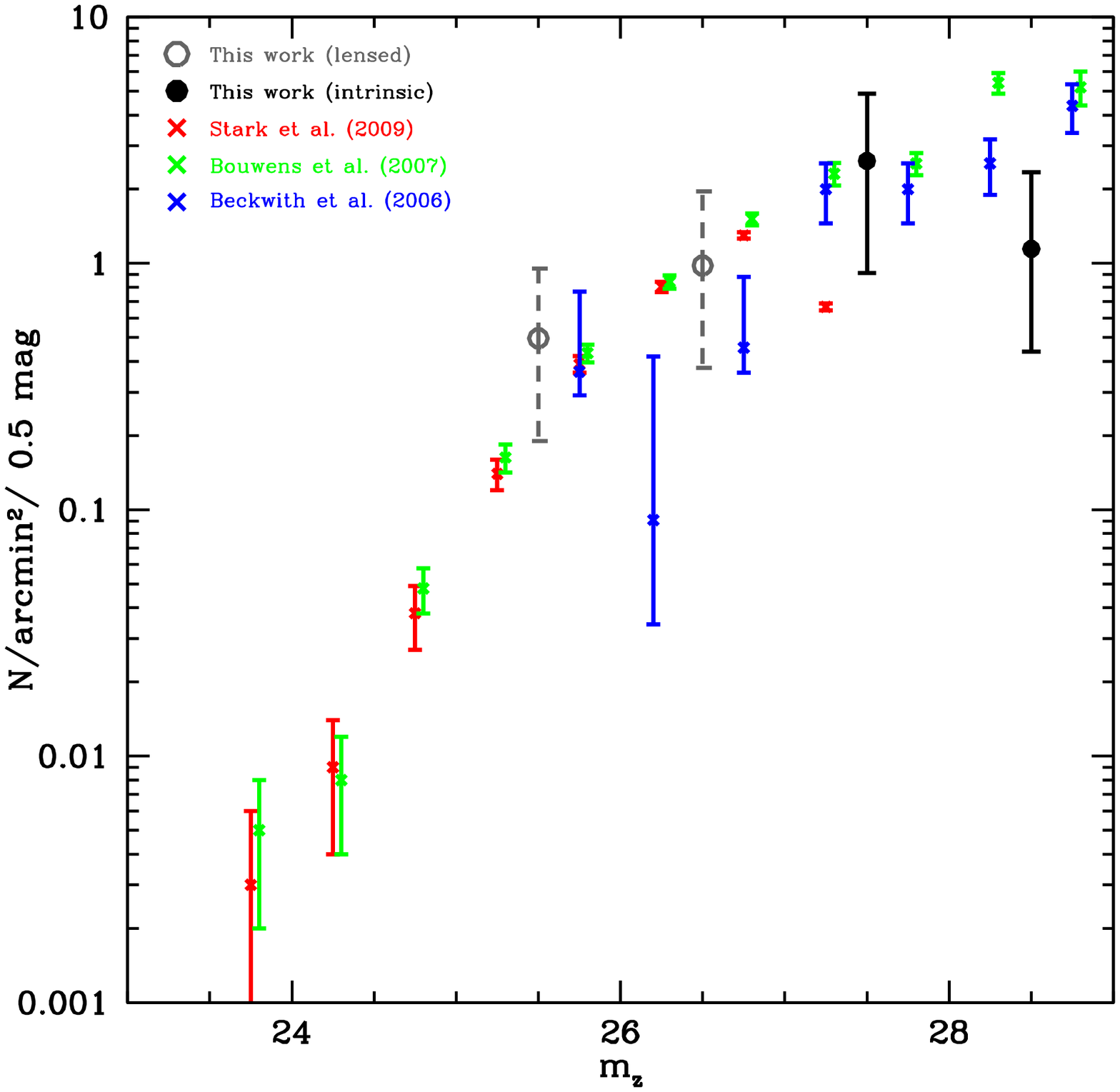}
\end{center}
\end{minipage}
\begin{minipage}{0.5\textwidth}
\begin{center}
\includegraphics[width=0.9\textwidth]{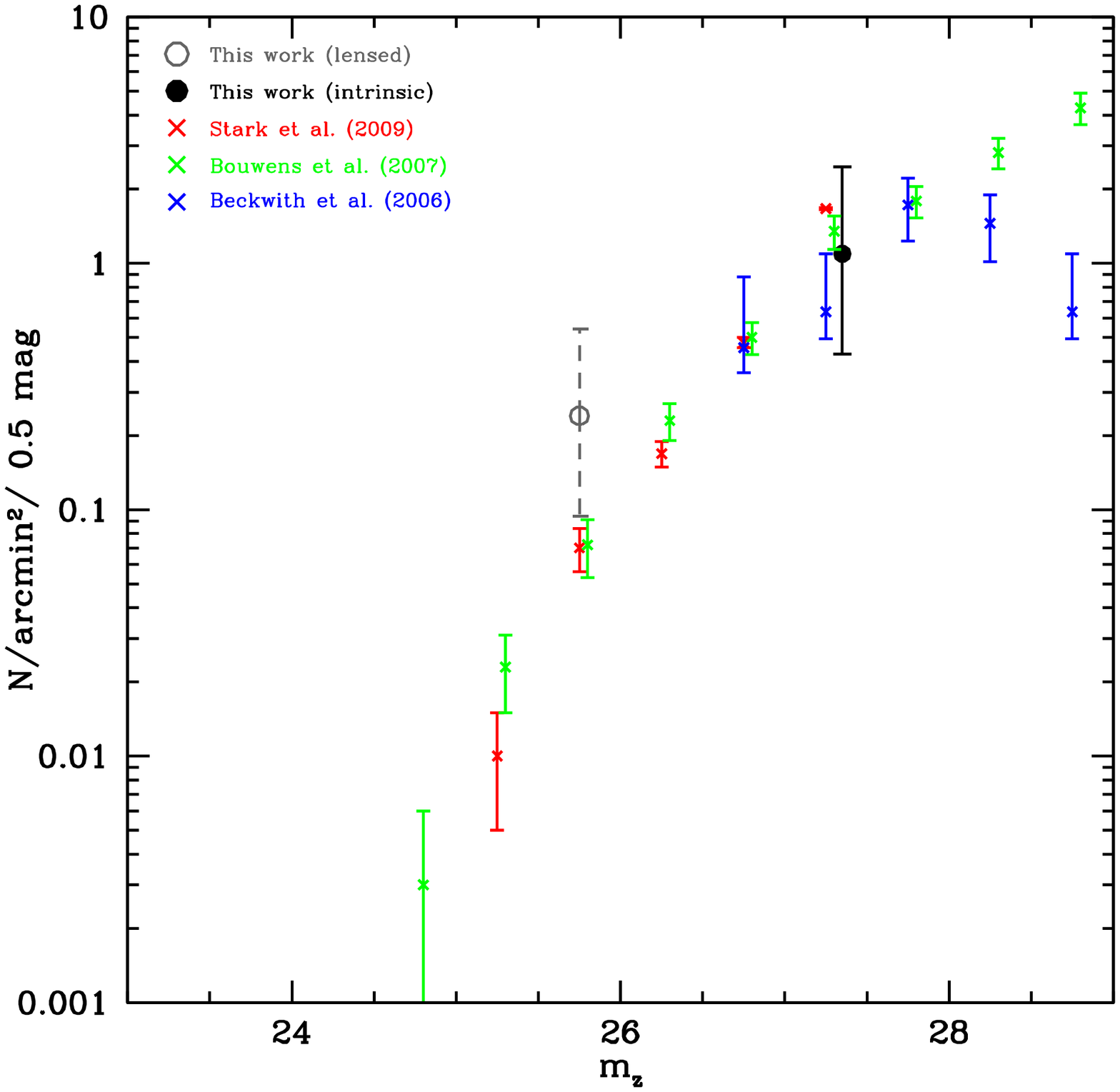}
\end{center}
\end{minipage}
\caption{Surface densities (number per $\mbox{arcmin}^2$ per
  $0.5\mbox{ mag}$) of $V$ (left) and $i$-band (right) dropouts
  (circles).   Due to small number statistics, we bin the data in two bins
    of $\magz = [25-26]$ and $[26-27]$ for $V$ and one bin $\magz =
    [25-26.5]$ for $i$-dropouts and recalculate them to $0.5$-mag
    bins, which is the binning used in the literature (crosses). Only
    sources for which our completeness is greater than 5\% (i.e. those
    with $\magz < 27$) are presented and completeness corrections are
    discussed in the text. Direct measurements (gray open circles
  with dashed errorbars) have been corrected for the effect of
  gravitational lensing, giving the intrinsic (unlensed) number counts
  (black full circles with solid errorbars). The errors for our data
  include sample variance, Poisson errors, and errors due to lensing
  (the latter of which are small in comparison).  The crosses
    are all calculated for the bins indicated by \citet{stark09} data,
    the other two are shifted by 0.05 and 0.1~mag for
    clarity. \citet{bouwens07}, and \citet{stark09} data have been
    corrected for completeness, whereas \citet{beckwith06} data were
    not. The \citet{stark09} points were calculated using GOODS data,
    \citet{beckwith06} only include HUDF number counts, whereas
    \citet{bouwens07} use data from HUDF, HUDF-Ps, and GOODS.}
\label{fig:numcount}
\end{figure*}

\begin{figure}[ht]
\begin{center}
\includegraphics[width=0.5\textwidth]{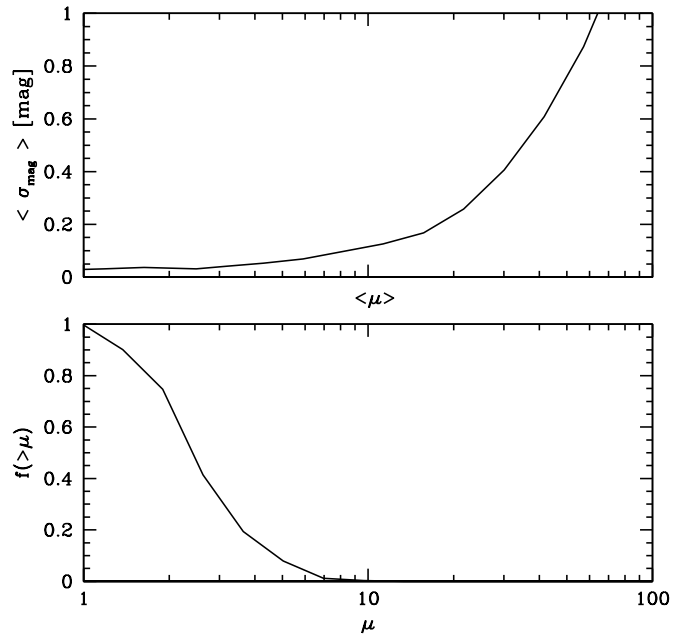}
\end{center}
\caption{Top: the average error on magnitude $\ave{\sigma_{\rm mag}}$
  as a function of magnification.  For magnifications of $\mu<10$, the
  errors are $<0.1\mbox{ mag}$ on average. The exact error as a
  function of location is given in
  \protect{Fig.~\ref{fig:errmap}}. Bottom: cumulative number of image
  plane pixels having magnifications larger than a certain
  value. $<1\%$ of the pixels will have $\mu>10$, which is the only
  regime where the errors are of any significance for determining the
  intrinsic luminosity of a sample.  Both plots assume a source at
    $z\to\infty$.}
\label{fig:magstats}
\end{figure}

In addition, the average error $\ave{\sigma_{\rm mag}}$ as a function
of average magnification $\ave{\mu}$ is shown in
Fig.~\ref{fig:magstats}.  As a comparison we also give the cumulative
area as a function of magnification $\mu$, this time calculated for
our reconstructed magnification map from Section~\ref{sec:data}.
These plots illustrate that less than $0.1\mbox{ arcmin}^2$ of the
survey (i.e. $<1\%$ of the area for a single pointing ACS observation)
 has magnifications $\mu>10$ and on average low errors. It is
true, that we are more likely to find images in regions of
large magnification; however only 3 out of 20 systems have highly
uncertain magnifications (see Table~\ref{tab:dropouts}). Hence the
errors on the positions of the magnitude bins in evaluating the number
counts (Fig.~\ref{fig:numcount}) are much smaller than the bin widths.

 Above investigations use a fixed smooth model to determine
  uncertainties. To further evaluate uncertainties due to profile
  uncertainties, we have investigated magnification properties of
  three very different smooth profiles, singular isothermal sphere
  (SIS), non-singular isothermal sphere (NIS) and Navarro Frenk and
  White (NFW) profile. NFW and NIS profiles give rise to two critical
  curves, radial and tangential, whereas SIS's radial critical
  curve shrinks to zero size. The existence of radial arcs and
  simulations tell us that SIS is not a valid description for cluster
  mass distribution. Systems with $\ge2$ multiply imaged systems with
  known redshifts (like the one used here), and even more so, with
  included weak lensing information can distinguish between these
  profiles (see e.g. \citealp{sand08, limousin08}).

Still, we have investigated their magnification properties. First we
match the Einstein radii of all three models, which is easily measured
in practice even with a single multiply imaged system with known
redshift. If also matching the radial critical curve, the change in
area (due to magnification) is $<1\%$. If only the tangential critical
curve is matched, and the radial ones differ by $\sim 10\mbox{
  arcsec}$ these changes are still below $<20\%$, only when we compare
SIS with NFW is the change in area significant. It is only significant
for small observing fields (like the one used here) and can be as high
as $50\%$, however this case is unrealistic as noted
before. Furthermore, for a field twice this size, the errors due to
uncertainties in smooth model are again below $5\%$. The individual
magnifications differ by $<20\%$ everywhere outside the tangential
critical curve and are mostly in $5\%$ regime, which is where most of
our sources are located.

In conclusion, the errors due to lensing are smaller than the sample
variance.  Taking into account sub-dominant errors, the increased number of
objects detected, and the ability to reach otherwise unobservably
faint magnitudes, it is advantageous to use well-modeled clusters as cosmic
telescopes to measure the average properties of background ($z\gtrsim
5$) objects, including their luminosity function. This conclusion is
not limited only to high-redshift sources.  The system K is at $z\sim
2.7$, but magnification factors of $\mu \sim 50$ make this ordinary
(i.e. not ultra-luminous) sub-mm source detectable with Spitzer/IRAC
\citep{gonzalez09}, AzTEC (Atacama Submillimeter Telescope Experiment,
\citealp{wilson08} and BLAST (Balloon-borne Large-Aperture
Submillimeter Telescope, \citealp{rex09}).

\begin{deluxetable*}{rrrrrrrr}
\tablecolumns{8}
\tablewidth{0pc}
\tablecaption{The properties of $V$-dropouts (top) and $i$-dropouts (bottom).}
\tablehead{\colhead{} & \colhead{Ra} &\colhead{Dec} & \colhead{$m_{\rm z} \pm \Delta_{\rm phot}$} & \colhead{$(V-i)$} &\colhead{$(i-z)$} &\colhead{$\mu$} & \colhead{$m_{\rm z,int} \pm \Delta_{\rm l+p}$}}  
\startdata
1 &$ 104.59377 $ &$ -55.970039 $ &$  26.74 \pm 0.07 $ &$   2.2 \pm 0.4 $   &$   0.69 \pm 0.14 $ &$   2.68 \pm 0.08      $&$  27.81^{+0.08}_{-0.08} $\\ 
2 &$ 104.60201 $ &$ -55.954700 $ &$  27.89 \pm 0.10 $ &$   1.8 \pm 0.2 $   &$  -2.42 \pm 0.15 $ &$   3.97 \pm 0.26      $&$  29.39^{+0.12}_{-0.12} $\\
3 &$ 104.60582 $ &$ -55.969944 $ &$  26.45 \pm 0.06 $ &$  >1.5           $ &$   0.20 \pm 0.11 $ &$   3.17 \pm 0.09      $&$  27.70^{+0.07}_{-0.07} $\\
4 &$ 104.61459 $ &$ -55.951771 $ &$  27.42 \pm 0.05 $ &$   2.2  \pm 0.1  $ &$  -0.54 \pm 0.10 $ &$   4.0 \pm 0.9        $&$  28.93^{+0.23}_{-0.28} $\\
5 &$ 104.61855 $ &$ -55.940559 $ &$  26.24 \pm 0.05 $ &$   3.3  \pm 0.5  $ &$   0.86 \pm 0.13 $ &$   5.8 \pm 1.2        $&$  28.15^{+0.21}_{-0.26} $\\
6 &$ 104.63613 $ &$ -55.925079 $ &$  25.53 \pm 0.05 $ &$   2.3 \pm 0.6 $   &$   0.59 \pm 0.10 $ &$   3.78 \pm 0.15      $&$  26.98^{+0.07}_{-0.07} $\\
7 &$ 104.64101 $ &$ -55.935631 $ &$  26.97 \pm 0.05 $ &$   2.3 \pm 0.7 $   &$   1.17 \pm 0.13 $ &$   6^{+15}_{-5}        $&$  29^{+2}_{-2} $        \\ 
8 &$ 104.64129 $ &$ -55.964703 $ &$  26.38 \pm 0.06 $ &$   2.3 \pm 0.2 $   &$   0.14 \pm 0.11 $ &$  12 \pm 2            $&$  29.09^{+0.18}_{-0.21} $\\
9 &$ 104.64133 $ &$ -55.966335 $ &$  25.04 \pm 0.05 $ &$   2.5 \pm 0.4 $   &$   0.97 \pm 0.11 $ &$   10 \pm 1           $&$  27.54^{+0.11}_{-0.12} $\\
10&$ 104.64437 $ &$ -55.973896 $ &$  25.81 \pm 0.05 $ &$   2.4 \pm 0.6 $   &$   0.87 \pm 0.10 $ &$   5.2 \pm 0.3        $&$  27.64^{+0.08}_{-0.08} $\\
11&$ 104.64603 $ &$ -55.969257 $ &$  25.68 \pm 0.05 $ &$   2.8 \pm 0.3 $   &$  -0.11 \pm 0.10 $ &$   6.5 \pm 0.8        $&$  27.70^{+0.14}_{-0.15} $\\
12&$ 104.65024 $ &$ -55.961578 $ &$  25.52 \pm 0.05 $ &$   5.7  \pm 4.0 $  &$   0.88 \pm 0.10 $ &$ 480 \pm 30           $&$  32.22^{+0.08}_{-0.09} $\\ 
13&$ 104.65026 $ &$ -55.966267 $ &$  25.53 \pm 0.05 $ &$   2.6 \pm 0.2 $   &$   0.18 \pm 0.10 $ &$  30_{-29}^{+1000}      $&$  29^{+4}_{-4} $       \\ 
14&$ 104.65306 $ &$ -55.930065 $ &$  26.30 \pm 0.05 $ &$   2.1 \pm 0.1 $   &$   0.12 \pm 0.10 $ &$   4.6 \pm 0.2        $&$  27.96^{+0.07}_{-0.07} $\\
15&$ 104.65884 $ &$ -55.948055 $ &$  27.23 \pm 0.06 $ &$   4.0 \pm 0.2 $   &$   0.06 \pm 0.11 $ &$ 207 \pm 3            $&$  33.02^{+0.06}_{-0.06} $\\
16&$ 104.66796 $ &$ -55.949810 $ &$  25.60 \pm 0.05 $ &$   3.2 \pm 0.6 $   &$   0.72 \pm 0.10 $ &$  34^{+100}_{-33}       $&$  29^{+2}_{-4} $       \\ 
17&$ 104.67584 $ &$ -55.945240 $ &$  26.01 \pm 0.05 $ &$   3    \pm 1.0  $ &$   0.32 \pm 0.10 $ &$   7 \pm 1.0          $&$  28.19^{+0.15}_{-0.17} $\\
18&$ 104.67677 $ &$ -55.943958 $ &$  28.18 \pm 0.05 $ &$   1.5 \pm 0.10 $  &$  -1.42 \pm 0.10 $ &$   7.1 \pm 0.5        $&$  30.31^{+0.09}_{-0.09} $\\
19&$ 104.68509 $ &$ -55.951725 $ &$  27.14 \pm 0.05 $ &$   3.9 \pm 0.2 $   &$  -0.23 \pm 0.10 $ &$   8.6 \pm 0.4        $&$  29.47^{+0.07}_{-0.07} $\\
20&$ 104.68786 $ &$ -55.943943 $ &$  27.36 \pm 0.09 $ &$   2.2 \pm 0.5 $   &$  -0.05 \pm 0.14 $ &$   4.8 \pm 0.2        $&$  29.06^{+0.10}_{-0.10} $\\\cline{1-8}
1&$ 104.63920 $ &$ -55.925922 $ &$  26.29 \pm 0.05 $ &$  >0.5 $ &$   1.33 \pm 0.10 $     &$   4.9 \pm 0.2  $&$  28.01_{-0.07}^{+0.07} $\\ 
2&$ 104.65271 $ &$ -55.930984 $ &$  25.44 \pm 0.05 $ &$  >1.4  $ &$   1.33 \pm 0.10 $    &$   4.6 \pm 0.2  $&$  27.10_{-0.07}^{+0.07} $\\ 
3&$ 104.66599 $ &$ -55.931286 $ &$  25.38 \pm 0.05 $ &$   2 \pm 1 $ &$   1.34 \pm 0.11 $ &$   3.4 \pm 0.2  $&$  26.71_{-0.08}^{+0.08} $\\ 
4&$ 104.67469 $ &$ -55.929668 $ &$  26.04 \pm 0.05 $ &$  >0.5  $ &$   1.71 \pm 0.13 $    &$   3.1 \pm 0.1  $&$  27.26_{-0.06}^{+0.06} $\\ 
\enddata 
\label{tab:dropouts}
\tablecomments{ For the final error on $m_{\rm z,int}$
    ($\Delta_{\rm l+p}$) photometric and magnification errors were added in quadrature.}
\end{deluxetable*}

\section{Conclusions and Outlook}
\label{sec:conclusions}
We have presented a new mass reconstruction for the post-collision
Bullet Cluster. In addition, we have used its extraordinary
capabilities as a cosmic telescope to study $z\sim 5-6$
objects. Our main conclusions are as follows:
\begin{enumerate}
\item Our new strong and weak lensing mass reconstruction method, that
uses a non-uniform adaptive pixel grid, performs very well in
reconstructing the cluster potential in the vicinity of multiple
images. In particular, we have improved the accuracy with which the
image positions are reproduced to an average rms residual of
$1.4\arcsec$. The limitation in pushing this number to
the observed uncertainties in image position ($\sim 0.1\arcsec$) is
mostly in the lack of secure redshifts (only one spectroscopic
redshift is known for the 12 multiple image systems used here). The
method can resolve the small scale
substructure and reconstruct the images even more precisely, once the
redshifts become available.
\item Bullet Cluster is an efficient cosmic telescope. We have used its
  power to search for $z\sim 5-6$ objects using the dropout
  technique. After correcting for the lensing effect and taking into
  account the errors from the imperfect magnification map, we derive
  the surface densities of $V$ and $i$-band dropouts. They match well
  the densities (in absolute value) derived from previous studies
  \citep{stark09,bouwens07,beckwith06} using either deeper or larger
  surveys (HUDF and GOODS). E.g. GOODS (release 1) data has similar
  exposure times ($5000\mbox{s}$ in $V$ and $i$ and $10660$ in $z$)
  over $360\mbox{ arcmin}^2$ field. Scaling total numbers of dropouts
  from \citet{stark09} (using area ratio between our and their survey)
  we would obtain 14 $V$ and 3.5 $i$-band
  dropouts; whereas we observe 20 and 4. In addition they are at
  fainter (intrinsic) magnitudes as their sample.
\item Our sample numbers are much smaller (as our search area is
  $10\mbox{ arcmin}^2$, compared to the area used in
  e.g. \citet{stark09,bouwens07} with $>300\mbox{ arcmin}^2$), and so
  the errorbars are dominated by sample variance and Poisson
  noise. However, our results show, that at the same observed
  magnitudes, we gain in the number of sources observed compared to a
  blank field of the same size. In addition, we observe intrinsically
  fainter sources than would otherwise be possible in a blank field
  observed to similar depths. Finally, multiply imaged sources are
  easily discriminated from contaminants, as lens model roughly
  predicts their redshifts, and hence can be distinguished from
  $z\sim2$ dusty galaxy and cool stars.
\end{enumerate}

This analysis can and will be extended to higher redshift objects (z
and J-dropouts).  The main requirement for such studies is deep
optical and near-IR data, as well as high resolution magnification
maps such as the one presented in this paper.  These studies can also be
readily expanded to a larger sample of clusters. With a sample of
$\sim 20$ massive clusters, with well understood magnification
properties and observed to similar depths as the Bullet Cluster we can
increase the expected number counts and reduce the errors due to
sample variance and Poisson sampling to those of GOODS (see
e.g. \citealp{stark09}). However, due to lensing, we can push
observations to at least a magnitude deeper than GOODS and  only
  slightly shallower than HUDF (in much shorter observing time).
Finally, highly magnified images ($\mu\gtrsim 50$) potentially allow
for (otherwise prohibitive) spectroscopic follow up of these high
redshift galaxies.  Highly magnified objects will often sample
populations that would otherwise be unobservable by any practical
means.

\begin{acknowledgements}
  We would like to thank Massimo Stiavelli for many useful discussions
  and help with data reduction/acquisition. We would like to thank
  Alaina Henry for her comments on this manuscript and Maxim
  Markevitch, for many useful discussions and help with IRAC and
  Chandra data. We would like to thank Moses Marsh for his assistance
  during the early-stages of this project. Support for this work was
  provided by NASA through grant numbers HST-GO-10200, HST-GO-10863,
  and HST-GO-11099 from the Space Telescope Science Institute (STScI),
  which is operated by AURA, Inc., under NASA contract NAS 5-26555 and
  NNX08AD79G. MB acknowledges support from NASA through Hubble
  Fellowship grant \#~HST-HF-01206 awarded by the STScI. TT
  acknowledges support from the NSF through CAREER award NSF-0642621,
  by the Sloan Foundation through a Sloan Research Fellowship, and by
  the Packard Foundation through a Packard Fellowship. DA acknowledges
  support from the U.S. Department of Energy under contract number
  DE-AC02-76SF00515.  PJM acknowledges support from the TABASGO
  foundation in the form of a research fellowship.  This research has
  made use of data obtained from the Chandra Data Archive and software
  provided by the Chandra X-ray Center (CXC).
\end{acknowledgements}

\bibliography{/home/marusa/latex/inputs/bibliogr_clusters,/home/marusa/latex/inputs/bibliogr_gglensing,/home/marusa/latex/inputs/bibliogr,/home/marusa/latex/inputs/bibliogr_cv,/home/marusa/latex/inputs/bibliogr_highz}
\bibliographystyle{apj}

\end{document}